\documentclass[aps,prb,twocolumn,showpacs,superscriptaddress,floatfix]{revtex4-1}
\usepackage{dcolumn}
\usepackage{amsmath}
\usepackage{graphicx,epsfig,psfrag}
\usepackage{amssymb}
\usepackage{natbib}
\usepackage{url}
\usepackage[breaklinks=true]{hyperref}
\usepackage{mathtools}
\usepackage{subfigure}
\hypersetup{
        colorlinks=true,
}
\usepackage{bookmark}
\usepackage{epic}

\usepackage{txfonts}
\usepackage{color}
\newcommand{\hn}[1]{{\color{red}{#1}}}

\begin{document}
\title{Time-dependent numerical renormalization group method for multiple quenches: towards exact results for the long-time limit of thermodynamic observables and spectral functions}
\author{H. T. M. Nghiem}
\affiliation
{Peter Gr\"{u}nberg Institut and Institute for Advanced Simulation, 
Research Centre J\"ulich, 52425 J\"ulich, Germany}
\affiliation{
Thanh Tay Institute for Advanced Study, Thanh Tay University, 
1000 Hanoi, Vietnam}
\affiliation{
Phenikaa Research and Technology Institute, A\&A Green Phoenix Group,
1000 Hanoi, Vietnam}
\affiliation{
Advanced Institute for Science and Technology, Hanoi University of
Science and Technology, 10000 Hanoi, Vietnam}
\author{T. A. Costi}
\affiliation
{Peter Gr\"{u}nberg Institut and Institute for Advanced Simulation, 
Research Centre J\"ulich, 52425 J\"ulich, Germany}
\begin{abstract}
We develop an alternative time-dependent numerical renormalization group (TDNRG) formalism for multiple quenches and implement it to study the response of a quantum impurity system to a general pulse. 
Within this approach, we reduce the contribution of the NRG approximation to numerical errors in the time evolution of observables by a formulation {that avoids the use of the generalized overlap matrix elements} in our
previous multiple-quench TDNRG formalism [Nghiem {\em et al.,} Phys. Rev. B {\bf 89}, 075118 (2014); Phys. Rev. B {\bf 90}, 035129 (2014)]. 
We demonstrate that the formalism yields a smaller cumulative error in the trace of the projected density matrix as a function of time and a smaller discontinuity of local observables between quenches than in our previous approach. 
Moreover, by increasing the switch-on time, the time between the first and last quench of the discretized pulse, the long-time limit of observables systematically converges to its expected value
in the final state, i.e., the more adiabatic the switching, the more accurately is the long-time limit recovered. The present formalism
can be straightforwardly extended to infinite switch-on times. We show that this yields highly accurate results for the long-time limit 
of both thermodynamic observables and spectral functions, and overcomes the significant errors within the single quench formalism [Anders {\em et al.}, Phys. Rev. Lett. {\bf 95}, 196801 (2005); Nghiem {\em et al.}, Phys. Rev. Lett. {\bf 119}, 156601 (2017)]. 
This improvement provides a first step towards an accurate description of nonequilibrium steady states of quantum impurity systems, e.g., within the scattering states NRG approach [Anders, Phys. Rev. Lett. {\bf 101}, 066804 (2008)].
\end{abstract}

\date{\today}
\maketitle

\section{Introduction}\label{sec:intro}
The response of strongly correlated quantum impurity systems to quenches, pulses, static, and time-dependent fields 
remains a challenging theoretical topic of relevance to a number of fields, including low-energy ion-surface scattering \cite{Langreth1991,Pamperin2015}, time dependent
dynamics and pumping in quantum dots \cite{Jauho1994,Sela2006,Schiller2008,Splettstoesser2005},  pump-probe spectroscopies of correlated electron materials \cite{Perfetti2006,Freericks2006,Freericks2009,Aoki2014}, 
and to proposed cold atom realizations of Anderson and Kondo impurity models \cite{Bauer2013,Nishida2013,Nishida2016,
Riegger2018,KanaszNagy2018} which may be probed in real time \cite{Cetina2016}. 

Techniques currently being used to investigate the time-dependent dynamics
of quantum impurity systems, include functional and real-time renormalization group methods \cite{Metzner2012,Kennes2012a,Schoeller2009}, flow equation \cite{Lobaskin2005,Wang2010}, quantum Monte Carlo \cite{Cohen2014a,Gull2011b,Weiss2008,Muehlbacher2008}, 
and density matrix renormalization group methods \cite{Daley2004,White2004,Schmitteckert2010}, the hierarchical quantum master equation approach \cite{Schinabeck2016,Schinabeck2018},  
and the time-dependent numerical renormalization group (TDNRG) method \cite{Anders2005,Anders2006,Anders2008a,Anders2008b,Eidelstein2012,Guettge2013,Nghiem2014a,Nghiem2014b,Nghiem2016,Nghiem2017}.  
However, no single technique is able to address in a nonperturbative and numerically exact way the time-dependent and nonequilibrium dynamics of quantum impurity systems in the interesting 
low-temperature strong-coupling regime.  For example, quantum Monte Carlo approaches become numerically expensive in the zero-temperature limit \cite{Gull2011b}; 
the functional renormalization group approach, while versatile, is often only quantitatively accurate for weak to intermediate interaction strengths \cite{Metzner2012,Khedri2017a};and the (single-quench) TDNRG approach
suffers from imperfect thermalization and finite errors in the long-time limit of observables due, primarily, to the logarithmic discretization of the bath inherent to this approach \cite{Rosch2012,Anders2006,Anders2008a,Anders2008b,Eidelstein2012,Nghiem2014a,Nghiem2014b,Nghiem2016,Nghiem2017}. 
Nevertheless, an approach for the response of quantum impurity systems to time-dependent fields based on the latter technique remains promising since it automatically builds in
the nonperturbative element of Wilson's (equilibrium) numerical renormalization group method \cite{Wilson1975,KWW1980a,Gonzalez-Buxton1998,Bulla2008}. Such an approach would therefore 
be highly suitable for accessing the low-temperature strong-coupling physics of quantum impurity models. 

In previous work \cite{Nghiem2014a,Nghiem2014b}, we proposed to improve the long-time limit of thermodynamic observables, following a switch from an arbitrary initial state to an arbitrary final state, within the TDNRG approach,
by replacing a single quench by a sequence of $n$ smaller quenches acting over a finite time $\tilde{\tau}_n$ (the switch-on time) within a multiple-quench generalization of the single-quench TDNRG approach. Within such an approach, which also generalizes the TDNRG approach to general pulses and periodic driving, we showed that the long-time limit of thermodynamic observables could be systematically improved by increasing the number of quenches and the switch-on time. This resulted in a significant improvement for the long-time limit of observables over the corresponding single-quench TDNRG results \cite{Nghiem2014a,Nghiem2014b}. Despite this improvement, the approach still suffers from a number of problems, which we outline below, and which, to a large extent, we overcome in this paper, 
in which we develop an alternative formulation of the multiple-quench TDNRG approach.

We identify several problems in the time evolution of observables as calculated within our previous multiple-quench TDNRG formalism for general pulses \cite{Nghiem2014a,Nghiem2014b}
(see Secs.~\ref{sec:formalism} and \ref{sec:accuracy} for more details): 
(i) the trace of the projected density matrix was found to deviate increasingly away from $1$ with increasing switch-on time $\tilde{\tau}_n$, the time required to switch from the initial to the final state, up to some finite switch-on time, before decreasing again for longer switch-on times (with a maximum deviation, however, below 1\%); (ii) the time evolution of an observable exhibited small discontinuities at the times corresponding to all but the first quench; and (iii) there was no easy way, within this formalism, to extract, other than numerically, the limit of an infinite switch-on time $\tilde{\tau}_n\to\infty$. 
The problems (i) and (ii) stem from the way the NRG approximation is implemented within this approach and the same problems are also encountered within 
a hybrid TDNRG approach to periodic switching \cite{Eidelstein2012}. As for (iii), having a formulation which allows the limit  $\tilde{\tau}_n\to\infty$ to be taken analytically would be advantageous for the following reason: we found in the approach of Refs.~\onlinecite{Nghiem2014a,Nghiem2014b} that the dependence of the long-time limit of an observable $O(t\to\infty)$ on $\tilde{\tau}_n$ was, in general, non-monotonic. While $O(t\to\infty)$ eventually converged with increasing $\tilde{\tau}_n$ to its correct value in the final state for sufficiently large $\tilde{\tau}_n$, it was not {\it a priori} evident how large $\tilde{\tau}_n$ should be for convergence to be achieved. This problem is overcome by our alternative formulation, which allows the limit $\tilde{\tau}_n\to\infty$ to be taken analytically. In addition,
this formulation yields a much faster convergence of $O(t\to\infty)$ with increasing $\tilde{\tau}_n$, which is moreover monotonic [see Fig.~\ref{fig:compare2old-b}(b) in Sec.~\ref{sec:accuracy}].

In this paper, we present an alternative TDNRG formalism for multiple quenches which largely overcomes the above problems, i.e., (i) the trace of the projected density matrix versus  $\tilde{\tau}_n$ remains significantly closer to $1$ for all $\tilde{\tau}_n$ (Figs.~\ref{fig:compare2old} and \ref{fig:compare2old-b} in Sec. ~\ref{sec:accuracy}); (ii) observables exhibit  significantly smaller discontinuities after each quench [Fig.~\ref{fig:compare2old}(c) in Sec.~\ref{sec:accuracy}]; and (iii) the limit of infinite switch-on times can be taken analytically within this formalism and allows obtaining the long-time limit of thermodynamic observables with high accuracy (Sec.~\ref{sec:results}).

Extending this formalism to spectral functions, we also recover the expected long-time value of the spectral function in the equilibrium final state with high accuracy (see Fig.~\ref{fig:infinite} in Sec.~\ref{subsec:spectral-function}). Within a scattering states approach to nonequilibrium steady states \cite{Anders2008a,Anders2008b}, such
a calculation would allow the low-temperature nonequilibrium steady-state spectral function and conductance of interacting quantum dots to be calculated accurately for arbitrary bias and gate voltages. Besides the relevance of this for experiments on quantum dots \cite{Scott2009,Kretinin2011,Scott2013}, it would also go beyond the recent exact Fermi liquid approach which addresses only the low bias voltage regime (relative to the Kondo scale) \cite{Oguri2018,*Oguri2018Erratum1,Oguri2018b,Filippone2018} and could serve as a useful benchmark for other approaches \cite{Cohen2014a,Cohen2014b,Dorda2014,Dorda2015,Fugger2018}. 

The outline of the paper is as follows: In Sec. \ref{sec:formalism}, the alternative multiple quench TDNRG formalism is derived for finite switch-on times to reduce the effect of the NRG approximation by {avoiding the use of the generalized overlap matrix elements} 
in Refs.~\onlinecite{Nghiem2014a,Nghiem2014b}. The improvement is shown by comparing calculations from the two formalisms for the resonant level model and the Anderson impurity model in Sec.~\ref{sec:accuracy}. In Sec. \ref{sec:results}, 
the (straight-forward) extension to infinite switch-on times is derived from the formalism presented in Sec.~\ref{sec:formalism}.
Applications are made to the long-time limit of the spectral function in Sec.~\ref{subsec:spectral-function}, with additional supporting results in Appendix \ref{sec:reverse-quenches}, and to thermodynamic observables (occupation and double occupation) in Sec.~\ref{subsec:static-observables} with comparison of the results to the
their expected values in the equilibrium final state. In addition, we show in Sec.~\ref{subsec:static-observables-ed} results
for the occupation number of the resonant level model calculated within exact diagonalization (ED) for multiple quenches in the infinite switch-on time limit, which further support the conclusions made within the multiple quench TDNRG formalism. The TDNRG expression for the spectral function for multiple quenches, for a finite and infinite switch-on time, is derived in Appendix \ref{sec:spectra}. Results for the switch-on time dependence of the spectral function are presented in Appendix \ref{sec:finiteS}, while
Appendix~\ref{sec:lambda-dependence} discusses the effect of different discretization parameters on the error in the projected density matrices. 
Finally,  Appendix \ref{sec:ExD} presents the generalization of ED to study the time evolution of a system following multiple quenches with both finite and infinite switch-on times for the exactly solvable resonant level model.

\section{Multiple quench TDNRG for general pulses: alternative formalism}
\label{sec:formalism}

\begin{figure}[ht]
\centering 
\includegraphics[width=0.29\textwidth]{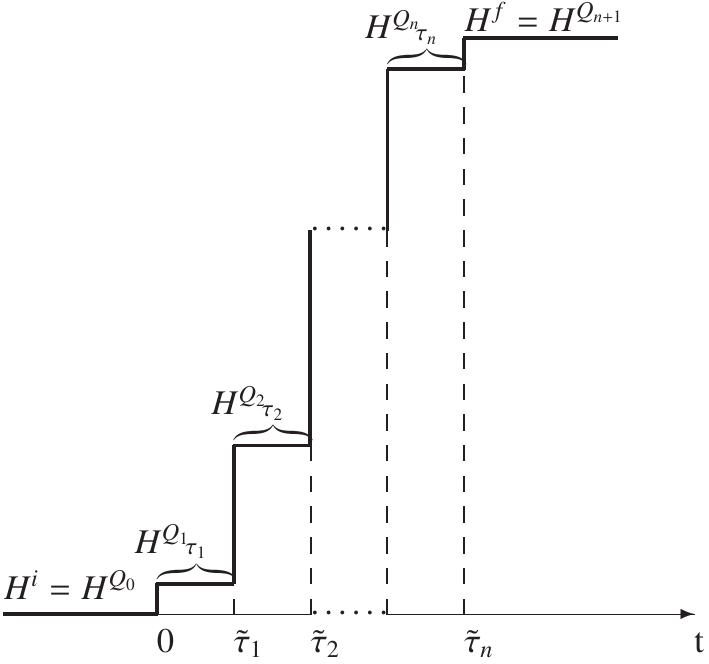}
\caption 
{A system driven from an initial to a final state via a sequence of quantum quenches at times $\tilde{\tau}_0=0,\tilde{\tau}_1,\dots,\tilde{\tau}_n$ with evolution according to
$\{H^{Q_p}\}$ in the time step $\tilde{\tau}_{p-1}\leq t < \tilde{\tau}_p$ with $H^{Q_0}=H^i$ and $H^{Q_{n+1}}=H^f$.
The time, $\tilde{\tau}_n=\sum_{p=1}^{n}\tau_p$, to switch  from $H^i$ to $H^f$ via the sequence of quantum quenches will be called 
the switch-on time (or, equivalently the pulse duration).
}
\label{fig:multiplesteps}
\end{figure}

We consider a system driven from an initial state (described by $H^{i}$) to a final state (described by $H^{f}$) in a time interval  $[0,\tilde{\tau}_n]$ 
via a sequence of $n+1$ quantum quenches described by $H^{Q_p},p=1,\dots,n+1$, switched on at times $\tilde{\tau}_{p-1},p=1,\dots,n+1$ (with $\tilde{\tau}_0=0$)
and having duration $\tau_p$ (except for  $H^{Q_{p=n+1}}=H^f$ which acts for all $t>\tilde{\tau}_n$) as depicted in Fig.~\ref{fig:multiplesteps}. The time to switch from the initial to the final
state, $\tilde{\tau}_n$, will be referred to as the switch-on time throughout the paper (equivalently, this can be called the duration of
the pulse).

The Hamiltonians, $H^{Q_p},p=1,\dots,n+1$, will represent an Anderson impurity model $H(t)$ for $\tilde{\tau}_{p-1}\leq t < \tilde{\tau}_p$, with
$$
H(t)= H_{\rm imp} + H_{\rm bath}+H_{\rm int}.
$$
Here, $H_{\rm imp}=\sum_{\sigma}\varepsilon_d(t)n_{d\sigma} + U(t)n_{d\uparrow}n_{d\downarrow}$ describes an impurity with a local level of energy $\varepsilon_d(t)$ and a Coulomb
repulsion $U(t)$ between opposite spin electrons in the local level.  The impurity interacts with free conduction electrons described by $H_{\rm bath}=\sum_{k\sigma}\epsilon_{k\sigma}c_{k\sigma}^{\dagger}c_{k\sigma}$ 
via a hybridization interaction $H_{\rm int}=V\sum_{k\sigma}(c_{k\sigma}^{\dagger}d_{\sigma}+h.c.)$. The time dependence enters through either a time dependent level position $\varepsilon_d(t)$ 
or a time dependent Coulomb repulsion $U(t)$ and will be specified in detail for each switching protocol later. We shall consider a time-independent hybridization $V$ throughout this paper,
and we shall denote the constant single-particle broadening of the resonant level by $\Gamma=\pi \rho V^2$, where $\rho=1/2D$ is the constant density of states per spin of the conduction electrons and $D=1$ is the
half bandwidth.

The quench Hamiltonians, $H^{Q_p},p=1,\dots,n+1$, are solved by using Wilson's NRG approach \cite{KWW1980a,Bulla2008} to yield the eigenstates and eigenvalues of each quench Hamiltonian
at each NRG iteration $m=m_{0},\dots,N$, where $N$ is the longest chain diagonalized and $m_{0}$ (typically 6 or 7) is the first iteration at which high energy states are discarded. 
In the iterative diagonalization of the Hamiltonians $H^{Q_p},p=1,\dots,n+1$ we retain either of order $N_s=1000$ states per 
iteration or truncate the spectrum at a fixed energy $E_{\rm cut}=24$ 
(measured in units of the characteristic scale $t_{m}\sim \Lambda^{-(m-1)/2}$ of the $m$-th truncated Hamiltonian   $H^{Q_p}_m$, where 
$\Lambda>1$ is the logarithmic discretization parameter \cite{Bulla2008}).
We make use of the complete basis set of discarded states\cite{Anders2005} $\{ |l_{p}e_{p}m_p\rangle_{Q_p}\}$ of $H^{Q_p}$ where $l_p$ labels the eigenstate,  $e_p$ the environment variable and $m_p$ the
truncated Hamiltonian at NRG iteration $m=m_p$ and the following decomposition of unity applies
$$
\sum_{l_{p}e_{p}m_{p}}  |l_{p}e_{p}m_p\rangle{_{Q_p}}  {_{Q_p}}\langle l_{p}e_{p}m_p| = 1.
 $$
In addition, in evaluating thermodynamic expectation values of observables, we  used the full density matrix representation \cite{Weichselbaum2007} for the initial state density matrix $\rho$ of $H^{i}$ and the z-averaging procedure \cite{Oliveira1994,Campo2005} to reduce discretization effects. 

With the above preliminaries, we can now write down the time evolution $O(t)$ of a local observable $\hat{O}$ at time $t\in [\tilde{\tau}_{p},\tilde{\tau}_{p+1})$. 
In the notation of Ref.~\onlinecite{Nghiem2014a,Nghiem2014b} we have
\begin{align}
O(t)= 
&\sum_{m_{p+1}l_{p+1}e_{p+1}}{_{Q_{p+1}}}\langle l_{p+1}e_{p+1}m_{p+1}|e^{-iH^{Q_{p+1}}(t-{\tilde\tau}_p)}e^{-iH^{Q_p}{\tau}_p}...e^{-iH^{Q_1}{\tau}_1} \nonumber\\
&\times \rho e^{iH^{Q_1}{\tau}_1}...e^{iH^{Q_p}{\tau}_p}e^{iH^{Q_{p+1}}(t-{\tilde\tau}_p)}\hat{O} |l_{p+1}e_{p+1}m_{p+1}\rangle{_{Q_{p+1}}}\label{eq:Otime},
\end{align}
with $\rho$ the full density matrix \cite{Weichselbaum2007} of the initial state Hamiltonian $H_{i}$ at inverse temperature $\beta=1/T$.
For the simplest case with $\tilde{\tau}_{1}>t\geq\tilde{\tau}_0$, the single quench result for $O(t)$ applies \cite{Anders2005}.
For the next simplest case with $\tilde{\tau}_{2}>t\geq\tilde{\tau}_1$, we have
\begin{align}
O(t)=&\sum_{m_{2}l_{2}e_{2}}{_{Q_{2}}}\langle l_{2}e_{2}m_{2}|e^{-iH^{Q_{2}}(t-{\tilde\tau}_1)}e^{-iH^{Q_1}{\tau}_1}\nonumber\\
&\times\rho e^{iH^{Q_1}{\tau}_1}e^{iH^{Q_{2}}(t-{\tilde\tau}_1)}\hat{O} |l_{2}e_{2}m_{2}\rangle{_{Q_{2}}}\nonumber\\
=&\sum_{m_{2}l_{2}e_{2}}\sum_{m'_{2}l'_{2}e'_{2}}\sum_{m_1l_1e_1}\sum_{m'_1l'_1e'_1}{_{Q_{2}}}\langle l_{2}e_{2}m_{2}|e^{-iH^{Q_{2}}(t-{\tilde\tau}_1)}|l_1e_1m_1\rangle{_{Q_{1}}}\nonumber\\
&\times{_{Q_{1}}}\langle l_1e_1m_1|e^{-iH^{Q_1}{\tau}_1}\rho e^{iH^{Q_1}{\tau}_1}|l_1'e_1'm_1'\rangle{_{Q_{1}}}\nonumber\\
&\times{_{Q_{1}}}\langle l_1'e_1'm_1'|e^{iH^{Q_{2}}(t-{\tilde\tau}_1)}|l'_{2}e'_{2}m'_{2}\rangle{_{Q_{2}}}{_{Q_{2}}}\langle l'_{2}e'_{2}m'_{2}|\hat{O} |l_{2}e_{2}m_{2}\rangle{_{Q_{2}}}\nonumber,
\end{align}
where three decompositions of unity $1=\sum_{lem}|lem\rangle\langle lem|$ have been employed.
Next, we use the identity \footnote{Equation (D4) in Weymann {\it et al.} \cite{Weymann2015}}
\begin{align}
\sum_{m_{2}l_{2}e_{2}}&\sum_{m'_{2}l'_{2}e'_{2}}\sum_{m_1l_1e_1}\sum_{m'_1l'_1e'_1}=\sum_m\sum_{e_1e_1'e_{2}e'_{2}}\sum_{r_1s_1r_{2}s_{2}}^{\notin K_1K_1'K_{2}K'_{2}}\label{eq:identity-multiple-shells}
\end{align}
to convert the multiple-shell sums over the four different Wilson chains in the above expression for $O(t)$ into 
a single shell-diagonal (restricted) sum involving kept states ($K_1K_1'$, etc. \cite{Anders2006}), obtaining
\begin{align}
O(t)=
\sum_m&\sum_{r_1s_1r_{2}s_{2}}^{\notin K_1K_1'K_{2}K'_{2}}\sum_{e_1e_1'e_{2}e'_{2}}{_{Q_{2}}}\langle r_{2}e_{2}m|e^{-iH^{Q_{2}}(t-{\tilde\tau}_1)}|r_1e_1m\rangle{_{Q_{1}}}\nonumber\\
&\times{_{Q_{1}}}\langle r_1e_1m|e^{-iH^{Q_1}{\tau}_1}\rho e^{iH^{Q_1}{\tau}_1}|s_1e_1'm\rangle{_{Q_{1}}}\nonumber\\
&\times{_{Q_{1}}}\langle s_1e_1'm|e^{iH^{Q_{2}}(t-{\tilde\tau}_1)}|s_{2}e'_{2}m\rangle{_{Q_{2}}}{_{Q_{2}}}\langle s_{2}e'_{2}m|\hat{O} |r_{2}e_{2}m\rangle{_{Q_{2}}}\nonumber\\
=
\sum_m&\sum_{r_1s_1r_{2}s_{2}}^{\notin K_1K_1'K_{2}K'_{2}}S_{r_{2}r_1}^m
\sum_{e}{_{Q_{1}}}\langle r_1em|\rho |s_1em\rangle{_{Q_{1}}}e^{-i(E^m_{r_1}-E^m_{s_1}){\tau}_1}\nonumber\\
&\times S^m_{s_1s_{2}} O^m_{s_{2}r_{2}}e^{-i(E^m_{r_{2}}-E^m_{s_{2}})(t-{\tilde\tau}_1)}\nonumber\\
=
\sum_m&\sum_{r_1s_1r_{2}s_{2}}^{\notin K_1K_1'K_{2}K'_{2}}S_{r_{2}r_1}^m
\rho^{i\to Q_1}_m(r_1,s_1)e^{-i(E^m_{r_1}-E^m_{s_1}){\tau}_1}\nonumber\\
&\times S^m_{s_1s_{2}} O^m_{s_{2}r_{2}}e^{-i(E^m_{r_{2}}-E^m_{s_{2}})(t-{\tilde\tau}_1)}\label{eq:Otime2}.
\end{align}
Here, {$S_{r_2r_1}^m$ is the overlap matrix element which is defined as $S_{r_2r_1}^m \delta_{e_2,e_1}={_{Q_{2}}}\langle r_{2}e_{2}m|r_1e_1m\rangle{_{Q_{1}}}$,  
$O^m_{s_{2}r_{2}}$ is the matrix elements of $\hat{O}$ that $O^m_{s_{2}r_{2}}\times\delta_{e'_{2},e_{2}}= {_{Q_{2}}}\langle s_{2}e'_{2}m|\hat{O} |r_{2}e_{2}m\rangle{_{Q_{2}}}$, 
and $\rho^{i\to Q_1}_m(r_1,s_1)=\sum_{e}{_{Q_{1}}}\langle r_1em|\rho |s_1em\rangle{_{Q_{1}}}$ is the reduced initial state density matrix (of $H_i$) projected onto the state of $H^{Q_1}$} \cite{Nghiem2014a}.
Furthermore, in the second line of Eq.~(\ref{eq:Otime2}), use has been made of the NRG approximation in the form $e^{iH^{Q_1}\tau_1}|r_1e_1m\rangle_{Q_1}\approx e^{iE^m_{r_1}\tau_1 }|r_1e_1m\rangle_{Q_1}$, which,
except in the limit of a vanishing switch-on time $\tau_1=0$, incurs a finite error in the time evolution, so Eq.~(\ref{eq:Otime2}) should be understood as being approximate. 

For the general case with $t\in [\tilde{\tau}_{p},\tilde{\tau}_{p+1})$,  we obtain, by using a generalization of Eq.~(\ref{eq:identity-multiple-shells})\cite{Note1},
\begin{align}
O(t)=\sum_m&\sum_{r_1s_1...r_ps_pr_{p+1}s_{p+1}}^{\notin K_1K_1'...K_pK_p'K_{p+1}K'_{p+1}}S_{r_{p+1}r_p}^m
...S_{r_2r_1}^m\rho^{i\to Q_1}_m(r_1,s_1)\nonumber\\
&\times e^{-i(E^m_{r_1}-E^m_{s_1}){\tau}_1}S^m_{s_1s_2}...e^{-i(E^m_{r_p}-E^m_{s_p}){\tau}_p}S^m_{s_ps_{p+1}}\nonumber\\
&\times O^m_{s_{p+1}r_{p+1}}e^{-i(E^m_{r_{p+1}}-E^m_{s_{p+1}})(t-{\tilde\tau}_p)}\label{eq:Otimep},
\end{align}
where, again, the use of the NRG approximation, implies that this expression should be understood, in general, as being approximate.
When $p=n$, Eq.~(\ref{eq:Otimep}) applies for all $t\geq \tilde{\tau}_n$, and can be used to extract the long-time limit $t\to\infty$ of observables, both for a finite  
or an infinite switch-on time $\tilde{\tau}_n$. Below, we shall discuss the accuracy of the long-time limit of observables $O(t\to\infty)$ as a function of the
switch-on time $\tilde{\tau}_n$ (or, equivalently the pulse duration). For zero switch-on time, $\tilde{\tau}_n=0$ (or equivalently $\tau_1=\tau_2=...=\tau_{p=n}=0$),  
the above expression can be converted into that for a single quench \cite{Nghiem2014a}.

For the special case that $\hat{O}$ is the identity operator, $\hat{O}=\hat{I}$, we have, \hn{using $O^m_{s_{p+1},r_{p+1}}=\langle s_{p+1}m|r_{p+1}m\rangle=\delta_{r_{p+1},s_{p+1}}$},
\begin{align}
1=\sum_m&\sum_{r_1s_1...r_ps_pr_{p+1}}^{\notin K_1K_1'...K_pK_p'K_{p+1}}S_{r_{p+1}r_p}^m
...S_{r_2r_1}^m\rho^{i\to Q_1}_m(r_1,s_1)\nonumber\\
&\times e^{-i(E^m_{r_1}-E^m_{s_1}){\tau}_1}S^m_{s_1s_2}...e^{-i(E^m_{r_p}-E^m_{s_p}){\tau}_p}S^m_{s_pr_{p+1}}\label{eq:Itimep}.
\end{align}
{This expression should, in general, be understood as approximate due to the use of the NRG approximation in its derivation. As a result,
the right hand side of this expression will deviate somewhat from 1 and will depend on time in a stepwise fashion through the condition 
$t\in [\tilde{\tau}_{p},\tilde{\tau}_{p+1})$.}
Equation (\ref{eq:Itimep}) is analogous to the trace  of the projected density matrix defined in Refs.~\onlinecite{Nghiem2014a,Nghiem2014b}, therefore the calculation of $\langle \hat{I}\rangle$ by using this 
equation will be referred to in the following as the trace of the projected density matrix, and will be denoted by $Tr[\rho^{i\to f}(t)]$ with $t\in [\tilde{\tau}_{p},\tilde{\tau}_{p+1})$. For $t>\tilde{\tau}_n$,
it is independent of time and denoted by $Tr[\rho^{i\to f}(\tilde{\tau}_n)]$. The deviation of $Tr[\rho^{i\to f}(\tilde{\tau}_n)]$ from 1 represents the cumulative error in the trace due to the NRG approximation   
and will be investigated in detail in the next section.  In the limit of a vanishing switch-on time, equivalent to a single quench, 
the NRG approximation is inoperative and the resulting expression $1=Tr[\rho^{i\to Q_1}]$ is satisfied exactly, as shown explicitly in Ref.~\onlinecite{Nghiem2014a}. 

Since we also wish to compare the present formalism with our previous multiple quench TDNRG formalism \cite{Nghiem2014a,Nghiem2014b},  a few words are in order about the latter. In Refs.~\onlinecite{Nghiem2014a,Nghiem2014b},
we expressed the time evolution of an observable $\hat{O}$ for $t\in [\tilde{\tau}_{p},\tilde{\tau}_{p+1})$  as
\begin{align}
O(t)=\sum_{mrs}^{\notin KK'}\rho^{i\to Q_{p+1}}_{rs}(m,\tilde{\tau}_p) e^{-i(E^m_r-E^m_s)(t-\tilde{\tau}_p)} O^m_{sr}\label{eq:Otime-old},
\end{align}
\begin{align}
 &\text{with}\quad\rho^{i\to Q_{p+1}}_{rs}(m,\tilde{\tau}_p)\nonumber\\
&= \sum_e{_{Q_{p+1}}}\langle rem|e^{-iH^{Q_p}{\tau}_p}\dots e^{-iH^{Q_1}{\tau}_1}\rho e^{iH^{Q_1}{\tau}_1}\dots e^{iH^{Q_p}{\tau}_p}| sem\rangle{_{Q_{p+1}}},\nonumber
\end{align}
a projected density matrix depending on each time step $\tilde{\tau}_p$ that was calculated recursively in terms of reduced density matrices and the 
so called {\em generalized overlap matrix elements} defined as
\begin{align}
\mathcal{S}^{m}_{r_is_{Q_{p+1}}}(\tilde{\tau}_p)\times\delta_{ee'}={_i}\langle rem|e^{iH^{Q_1}{\tau}_1}\dots e^{iH^{Q_p}{\tau}_p}|se'm\rangle{_{Q_{p+1}}}\label{eq:overlapTime}.
\end{align}
{These generalized overlap matrix elements are also calculated recursively via two recursion relations: (i) the matrix elements at $\tau_p$ are calculated recursively from the matrix 
elements of the previous time step at $\tau_{p-1}$, as shown in Eqs. (19) and (20) in Ref.~\onlinecite{Nghiem2014b}, and, (ii), 
the matrix elements of shell $m$ are also calculated recursively from the 
matrix elements of shell $(m-1)$ as in Eq.~(21) of Ref.~\onlinecite{Nghiem2014b}.  
Due to these recursion relations, the projected density matrix includes errors from
the NRG approximation not only from terms involving intra-shell excitations
$E^m_{r_q} - E^m_{s_q}$ but also from terms involving inter-shell excitations
$E^m_{r_q} - E^n_{s_q}$ with $n=m_0, m_0+1, ..., m-1$\footnote{Due to such inter-shell excitations appearing in the generalized overlap matrix elements, the eigenvalues $E^m_{r_q}$ and $E^n_{s_q}$ from different shells have to be measured relative to a common groundstate energy, chosen to be the groundstate of the longest Wilson chain. In this sense, absolute energies, measured relative to the latter groundstate, enter in the formalism of Ref.~\onlinecite{Nghiem2014b}}. 
In the present approach, by using a general form of Eq.~(\ref{eq:identity-multiple-shells}), we can 
derive Eq.~(\ref{eq:Otimep}) in which no recursion relation is needed (only ordinary overlap matrix elements $S_{r_{p+1}r_p}^m$ 
appear), and the projected density matrix includes only terms with intra-shell excitations 
$E^m_{r_q} - E^m_{s_q}$. Since the NRG eigenvalues are only approximations to the true eigenvalues, 
the projected density matrix in the previous approach includes more approximated terms than 
that in the present approach. For this reason, we expect, and find that the present approach is more 
accurate  than the previous approach. In addition, a recursive evaluation of the generalized overlap matrix 
elements of the previous approach is numerically more demanding than that of the ordinary overlap
matrix elements, so the present approach is also numerically more efficient and easier to implement 
than the approach of  Ref.~\onlinecite{Nghiem2014b}. 
}

Finally, within the multiple quench formalism of Ref.~\onlinecite{Nghiem2014a,Nghiem2014b}, the limit of
infinite switch-on time $\tilde{\tau}_n\to+\infty$ is impossible to take analytically, and that formalism is restricted to numerical evaluations at finite switch-on times. Within the present formalism, on the other hand, it becomes straightforward to take this limit (see Sec.~\ref{sec:results}). This, in turn, allows for an adiabatic switching of the system between an arbitrary initial state and an arbitrary final state, thereby improving the long-time limit of observables.

\section{Comparison with the previous approach}
\label{sec:accuracy}
\begin{figure}[h]
  \centering 
  \includegraphics[width=1\linewidth]{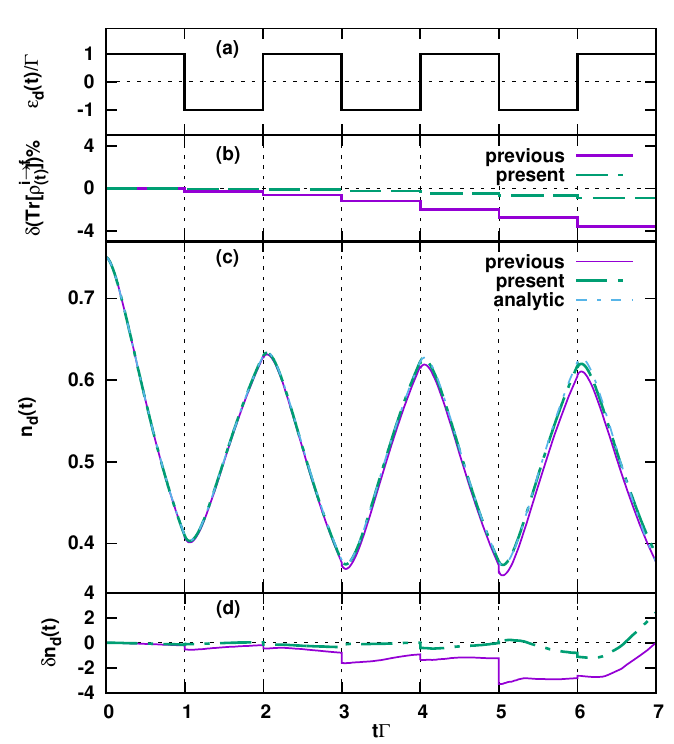} 
  \caption{
    Results for the resonant level model subject to square periodic driving.
    (a) The square periodic driving used for $\varepsilon_d(t)/\Gamma$, where $\varepsilon_d(t)$ is the local level position and $\Gamma$ the hybridization strength in the resonant level model.
    (b) Percentage deviation, $\delta({\rm Tr}[\rho^{i\to f}(t)])$, of the trace of the projected density matrix away from $1$ vs $t\Gamma$ in the present approach 
    (dashed line) and the previous multiple-quench approach of Refs.~\onlinecite{Nghiem2014b} (solid line). 
    (c) Occupation number $n_d(t)$ vs $t\Gamma$ in the present approach (dashed line), the previous approach (solid line), and, in the exact analytic approach (dash-dotted line).
    (d) Percentage deviation, $\delta n_d(t)$, of the occupation number vs  $t\Gamma$, from the exact analytic result, in the previous approach (solid line), and, in the 
    present approach (dashed-dotted line). 
    {NRG parameters for the TDNRG calculations: $\Lambda= 1.6$, $N_s=900$ kept states in each iteration, and $N_z=16$ values for the $z$-averaging.}
    \label{fig:compare2old}
}
\end{figure}
  
\begin{figure}[h]
    \centering 
  \includegraphics[width=1\linewidth]{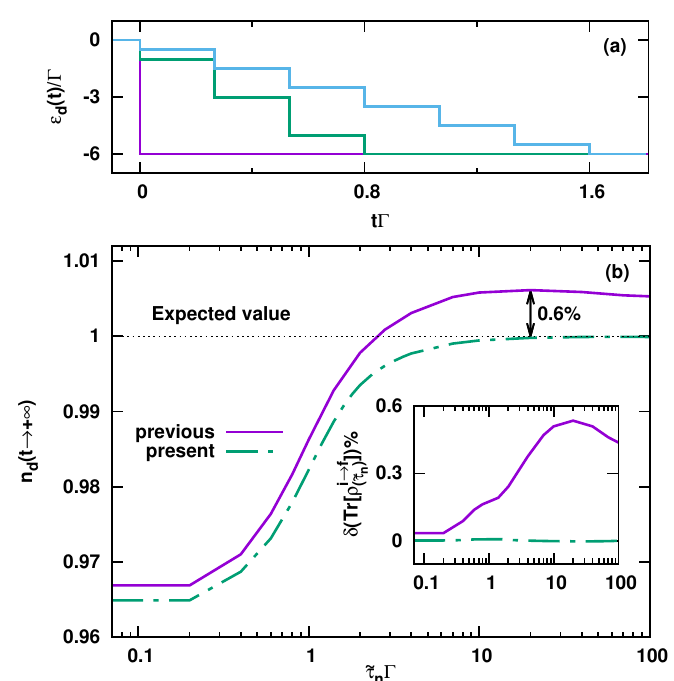}
  \caption{Anderson model subject to a linear ramp pulse.
    (a) A single large quench for $\varepsilon_d(t)/\Gamma$ is replaced by a linear ramp pulse and the latter is approximated by a finite sequence of $n>1$ small quenches of total duration $\tilde{\tau}_n$ (the switch-on time ).  
    The system is switched from the mixed valence regime with 
    $\varepsilon_d(t<0)=0$ and $U=12\Gamma$ to the symmetric Kondo regime with $\varepsilon_d(t\geq \tilde{\tau}_n)=-U/2$ and $U=12\Gamma$ within time $\tilde{\tau}_n$. 
    (b) Occupation number in the long-time limit $n_d(t\to\infty)$ vs $\tilde{\tau}_n\Gamma$ in the present approach (dashed-dotted line) compared to the approach of Ref.~\onlinecite{Nghiem2014b} (solid line). 
    For each fixed $\tilde{\tau}_n$, the linear ramp pulse is approximated by a sequence of up to 100 small quenches, with the number of quenches chosen such that $n_d(t\to\infty)$ is converged.
    The inset  shows the corresponding percentage error in the trace of the projected density matrix $\delta({\rm Tr}[\rho^{i\to f}(\tilde{\tau}_n)])$ vs $\tilde{\tau}_n\Gamma$ of the present (dot-dashed line) and previous (solid line) multiple-quench approach.
    NRG parameters: $\Lambda= 4$,  $E_{\text{cut}}=24$, and $N_z=8$ values for the z-averaging.  
}
  \label{fig:compare2old-b}
\end{figure}

In this section we illustrate the improvement of the present multiple-quench TDNRG approach over our previous approach for two specific situations: (i) for the time evolution of the occupation number $\langle n_d(t)\rangle$ in the resonant level model
\footnote{In contrast to the Anderson model, the RLM is a spinless model so within the TDNRG calculations for the latter one can retain a larger number of states than for the former.} under
 a square periodic driving of the local level and (ii) for the convergence of the long-time limit of the occupation number $\langle n_d(t\to\infty)\rangle$ with respect to increasing the switch-on time in the interacting Anderson impurity model following 
a linear ramp of the local level.

In Fig.~\ref{fig:compare2old}(b), we show the error in $Tr[\rho^{i\to f}(t)]$ versus time and the time evolution of the occupation number in the resonant level model (RLM) under a square periodic driving of the local level $\varepsilon_d$ from $-\Gamma$ to $\Gamma$ and back with a period of $2/\Gamma$ [Fig.~\ref{fig:compare2old}(a)]. The results of the present approach are compared with those from our previous multiple-quench formalism in Ref.~\onlinecite{Nghiem2014b} as well as with the exact analytic result for the RLM.
From these comparisons, we see that the previous formalism yields a trace for the projected density matrix ($Tr[\rho^{i\to f}(t)]$) which deviates increasingly away from $1$ after each quench. Similarly, the discontinuity in the time evolution of the occupation number at the boundaries of the time steps is clearly visible for times $t\Gamma\gtrsim 5$ in the results from the previous formalism. Within the present formalism, the deviation of $Tr[\rho^{i\to f}(t)]$ away from $1$ is reduced by a factor of more than $10$ relative to that in the previous formalism after one period, and the discontinuity in the time evolution of the occupation number also decreases by a similar factor. The present formalism results in a time evolution for $n_d$  which is significantly closer to the exact analytic one than that from the previous formalism, {as illustrated in Fig.~\ref{fig:compare2old}(c) and \ref{fig:compare2old}(d).}

Figure~\ref{fig:compare2old-b} shows results for the Anderson model with a constant Coulomb repulsion $U_i=U_f$ in which the system is switched from the mixed valence regime initially ($\varepsilon_{d}^i=0$) to the symmetric Kondo 
regime in the final state ($\varepsilon_d^f=-U_f/2$): in particular, we show the occupation number in the long-time limit $n_d(t\to\infty)$ and the corresponding percentage 
error in the trace of the projected density matrix as a function of the switch-on time $\tilde{\tau}_n$, comparing the results also with those from our previous approach. We see that $n_d(t\to\infty)$ initially increases as the switch-on time increases in both approaches. However, while the occupation number in the previous approach eventually overshoots the expected value of $1$ in the final state and only begins to drop close to the correct value at very long switch-on times, 
the present approach converges monotonically to the correct value already at relatively short switch-on times without overshooting [Fig.~\ref{fig:compare2old-b}(b)]. 
The difference to the expected value at the longest switch-on time $\tilde{\tau}_n\Gamma=100$ is less than $10^{-4}$ for the present improved approach. 
This significant improvement is also observed for the cumulative ($t=\infty>\tilde{\tau}_n$) error in $Tr[\rho^{i\to f}(\tilde{\tau}_n)]$. While this  is at most $\sim 0.6\%$ in the previous approach, 
the present formalism yields a value of less than $0.01\%$ in the whole range of switch-on times [see inset to Fig.~\ref{fig:compare2old-b}(b)].

In general, then, the present formalism for multiple quenches results in an improved time evolution for observables, including an improved long-time limit of observables and smaller discontinuities of observables after each quench. 
In the next section, we present and discuss the extension of this formalism to  strictly infinite switch-on times.

\section{Infinite switch-on time and accurate results in the long-time limit}
\label{sec:results}

In this section, we extend the formalism in Sec.~\ref{sec:formalism} to the infinite switch-on time limit and apply this to the long-time limit of the spectral function and local thermodynamic observables in
the interacting Anderson impurity model. We show that the resulting long-time limit of the spectral function (Sec.~\ref{subsec:spectral-function}) and local thermodynamic observables
 (Sec.~\ref{subsec:static-observables}) approach their expected values in the equilibrium final state to high accuracy. This conclusion is further supported by a (multiple quench) exact diagonalization 
study of the local level occupation number in the resonant level model (Sec.~\ref{subsec:static-observables-ed}).

The limit of an infinite switch-on time,  $\tilde{\tau}_{n}\to +\infty$, can be implemented in Eq.~(\ref{eq:Otimep})  by applying the restriction that 
{$r_1=s_1,r_2=s_2,...,r_n=s_n$,}
 resulting in
\begin{align}
O(t>\tilde{\tau}_n\to\infty)=\sum_m&\sum_{r_1...r_nr_{n+1}s_{n+1}}^{\notin K_1...K_nK_{n+1}K'_{n+1}}S_{r_{n+1}r_n}^m
...S_{r_2r_1}^m\rho^{i\to Q_1}_m(r_1,r_1)\nonumber\\
\times&S^m_{r_1r_2}...S^m_{r_ns_{n+1}} O^m_{s_{n+1}r_{n+1}}e^{-i(E^m_{r_{n+1}}-E^m_{s_{n+1}})(t-{\tilde\tau}_n)}\label{eq:Otimep_inf},
\end{align}
where in the above $t-\tilde{\tau}_{n}$ may still be finite.
In the long-time limit, infinitely long after the last quench, $O(t-\tilde{\tau}_n\to\infty)$ is calculated by applying the restriction that $r_{n+1}=s_{n+1}$ to the above equation. 

\begin{figure}[t]
  \centering
  \includegraphics[width=0.5\textwidth]{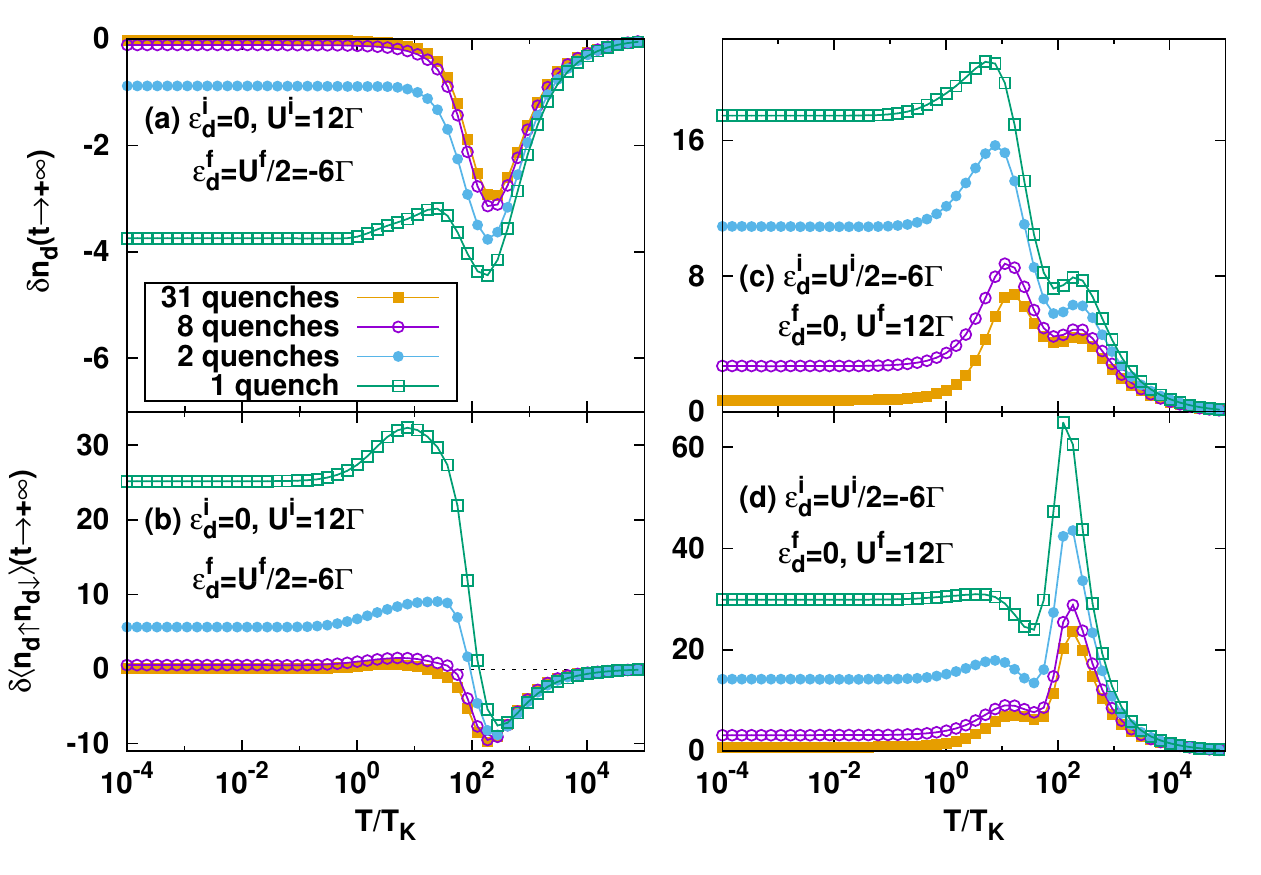}
    \caption{The percentage error in the expectation value of local observables in the long-time limit vs rescaled temperature $T/T_{\rm K}$ after a switch from the mixed valence to symmetric Kondo regime, (a) and (b), and a switch from the symmetric Kondo regime to the mixed valence regime, (c) and (d), with initial and final state parameters shown in the legends. The system is switched by applying $1$, $2$, $8$, or $31$ quenches on the local level and/or local Coulomb term, either, only on $\varepsilon_d$ in (a) and (b), or, on both $\varepsilon_d$ and $U$ in (c) and (d). (a) and (c) show the errors of the occupation numbers in the long-time limit, (b) and (d) show the errors in the double occupancy. $T_{\rm K}$ is the Kondo temperature of the symmetric system, $\Gamma=10^{-3}D$, and $D=1$ is the half-bandwidth. The calculations are for $\Lambda=4.0$, $E_{\text{cut}}=24$ , and $N_z=4$ values were used for the z-averaging. The results are normalized by the numerically calculated trace of the projected density matrix at each temperature.}  \label{fig:static}
\end{figure}

{A few remarks are in order concerning the implementation of the infinite switch-on time limit and the infinite time limits in the
above expression. In the limit of an infinite switch-on time, only the non-oscillatory part of terms such as $\lim_{\tau_p\to\infty}e^{i(E^m_{r_p}-E^m_{s_p})\tau_p}$ in Eq.~(\ref{eq:Otimep}) are finite, and yield $\delta_{E^m_{r_p},E^m_{s_p}}$(see Ref.~\onlinecite{Anders2006}). In the absence of degeneracies, we then have $\delta_{E^m_{r_p},E^m_{s_p}}=\delta_{r_p,s_p}$, i.e., the restriction $r_p=s_p$ used in Eq. (\ref{eq:Otimep_inf}). Since we implemented the U(1) charge and SU(2) spin symmetries explicitly for the Anderson model calculations in this work, all degeneracies are correctly taken into account. In general, however, when fewer symmetries are implemented, or when additional degeneracies arise during the renormalization group flow, conditions such as $r_p=s_p$ in our expressions for the infinite switch-on time limit should be replaced by $\delta_{E^m_{r_p},E^m_{s_p}}$. The latter equal energy condition is then implemented, in practice, by considering contributions from all states such that 
$|E^m_{r_p}-E^m_{s_p}|/t_m\ll 1$, with $t_m$ the low energy scale at iteration $m$. 
The same considerations apply to the long-time limit $t-\tilde{\tau}_n\to\infty$ of Eq.~(\ref{eq:Otimep_inf}). 
In specific cases, such as for the results in Figs.~\ref{fig:static} and Fig.~\ref{fig:IRLM}, we explicitly 
verified that both the above ways of implementing the equal energy restriction gave results for the 
long-time limit of observables lying within $10^{-8}$ of each other at all temperatures. 
}

Similarly, {we have from Eq.~(\ref{eq:Itimep})} for the trace of the projected density matrix in the limit ${\tilde{\tau}}_p\to +\infty$ {with $t\in [\tilde{\tau}_{p},\tilde{\tau}_{p+1})$},
\begin{align}
I=\sum_m&\sum_{r_1...r_pr_{p+1}}^{\notin K_1...K_pK_{p+1}}S_{r_{p+1}r_p}^m
...S_{r_2r_1}^m\rho^{i\to Q_1}_m(r_1,r_1)S^m_{r_1r_2}...S^m_{r_pr_{p+1}} \label{eq:Itimep_inf}.
\end{align}
This equality is not satisfied exactly due to the use of the NRG approximation inherent in its derivation, but as demonstrated in Sec.~\ref{sec:accuracy}, the deviation of the trace from 1 is small.
The small error is another reflection of the error in the long-time limit of an observable within TDNRG. 

\subsection{Application to the long-time limit of the spectral function}
\label{subsec:spectral-function}

The general expression for the time-dependent local spectral function $A(\omega,t)$ of the Anderson impurity model for times after the pulse (i.e., for $t>\tilde{\tau}_n$) within the present multiple quench TDNRG approach is
derived in Appendix~\ref{sec:spectra}. We use this here in the limit   $\tilde{\tau}_n\to\infty$ to discuss the long-time limit of the spectral function $A(\omega)=A(\omega,t\to\infty)$.

\begin{figure}[ht]
  \centering
  \includegraphics[width=0.5\textwidth]{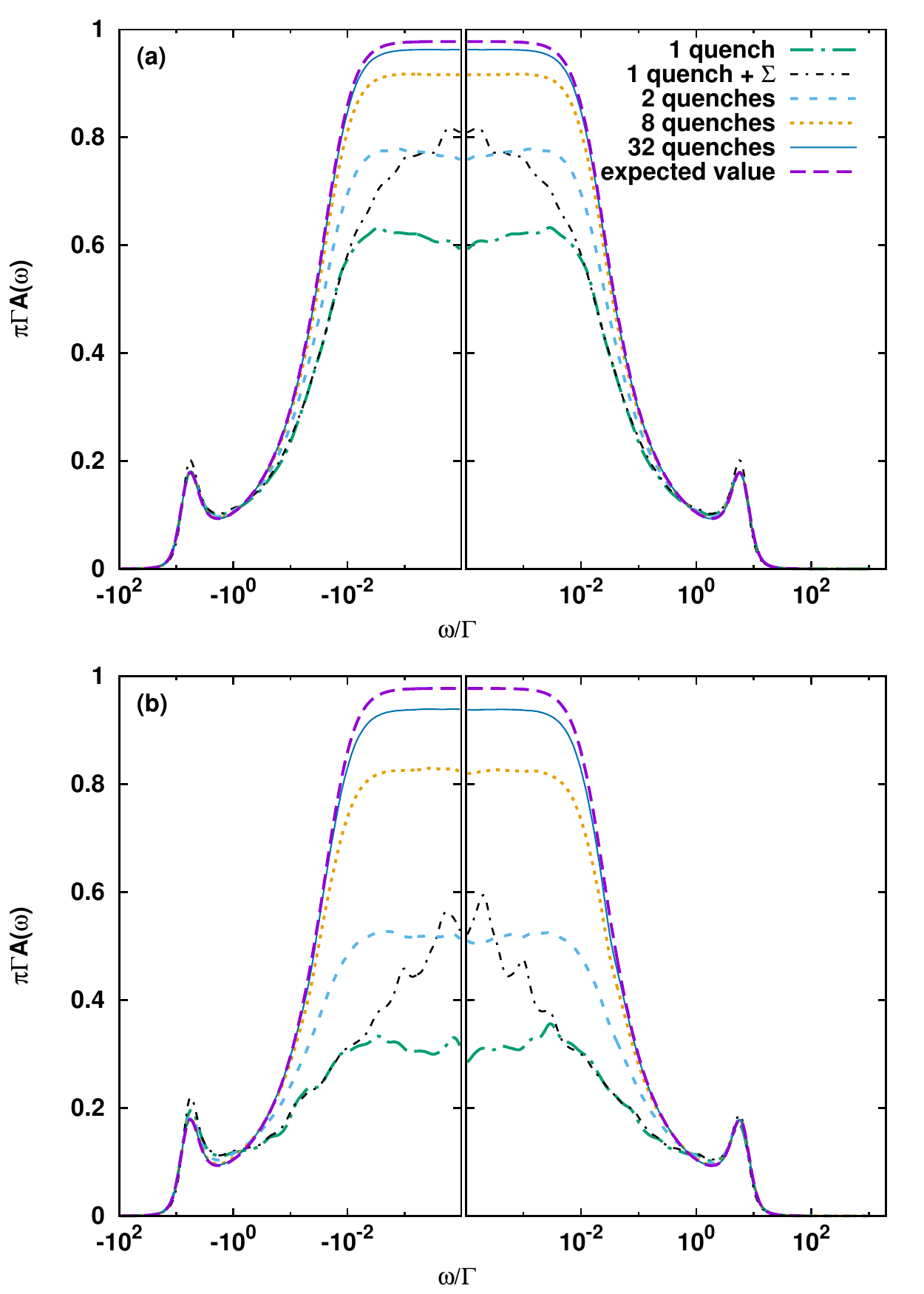}
\caption{
Normalized spectral function, $\pi\Gamma A(\omega)$, vs normalized frequency, $\omega/\Gamma$, in the long-time limit and infinite switch-on time for $2$, $8$, and $32$ quenches 
compared to that from the single-quench TDNRG with (1 quench $+\Sigma$) and without (1 quench) 
the use of the self-energy \cite{Bulla1998}.  Also shown is the expected value of the spectral function in the equilibrium final state. 
 (a): for switching from the noninteracting with $\varepsilon_d^i=U^i=0$ to the interacting system with $\varepsilon_{d}^f=-U^f/2, U^f=12\Gamma$. 
(b):  for switching from the mixed valence regime with $\varepsilon_d^i=0$ and $U^i=12\Gamma$ to the symmetric Kondo regime with $\varepsilon_{d}^f=-U^f/2, U^f=12\Gamma$. 
$\Gamma=10^{-3}D$, and $D=1$ is the half-bandwidth. Calculations were for essentially zero temperature $T=10^{-4}T_{\rm K}$, with $T_{\rm K}$ the Kondo temperature in the final state, 
NRG parameters: $\Lambda= 4$,  $E_{\text{cut}}=24$, and $N_z=8$ values for the z-averaging.
}   
\label{fig:infinite}
\end{figure}

Figures~\ref{fig:infinite}(a) and \ref{fig:infinite}(b) show  $A(\omega)$  for a system that is gradually driven, (a), 
from an uncorrelated symmetric initial state to a correlated symmetric Kondo regime, and, (b), 
from a mixed valence regime to the symmetric Kondo regime. We use a logarithmic energy axis to focus attention on the long-time limit of the low energy Kondo resonance at $|\omega|\lesssim T_{\rm K}$.  For both switching protocols, 
we show results for 1, 2, 8, and 32 quenches and also the results expected in the equilibrium final state and the single-quench result obtained by using the correlation self-energy $\Sigma$ to improve the calculation of $A(\omega)$ \cite{Bulla1998}. 
In Fig.~\ref{fig:infinite}(a), the single quench result without the use of the correlation self-energy has a Kondo resonance which achieves only 60\% of its Friedel sum rule value of 1 at $\omega=0$ \footnote{The Friedel sum rule for the Anderson impurity model 
states that $\pi\Gamma A(\omega=0)=\sin^2(\pi n_d/2)$ where $A(\omega)$ is the zero temperature equilibrium spectral function\cite{Hewson1997}. For a particle-hole symmetric final state, the right-hand-side equals $1$.}, while the improvement in the single quench result upon using 
the correlation self-energy to calculate $A(\omega)$ is not sufficient to reduce the error in the Friedel sum rule to below 20\%. In addition, the single quench TDNRG result for spectral functions suffer from additional substructures within the
Kondo resonance at $|\omega|\lesssim T_{\rm K}$, noticeable in Fig.~\ref{fig:infinite}(a), and discussed in detail elsewhere \cite{Nghiem2017}. On the other hand, a real improvement in the low energy Kondo resonance is observed within the multiple quench formalism
upon increasing the number of quenches, with eight quenches already yielding acceptable spectral functions with a less than 10\% error in the Friedel sum rule and with 32 quenches yielding highly accurate results approaching the expected value of the spectral function in the equilibrium final state. The substructures are also absent for this number of quenches.

Similar conclusions also hold for the second type of switching shown in Fig.~\ref{fig:infinite}(b), in which the system is switched from the mixed valence to the symmetric Kondo regime. 
While signatures of the initial state particle-hole asymmetry in $A(\omega)$ are present in the final state spectral function for the 1, 2, and 8 quench results, this asymmetry
is eliminated after 32 quenches, restoring the correct symmetry of the final state spectral function, which again also recovers accurately the expected equilibrium spectral function in the final state. 
In Appendix~\ref{sec:reverse-quenches} we also consider the reverse of the quenches shown in Figs.~\ref{fig:infinite}(a) and \ref{fig:infinite}(b), i.e., from a correlated to an uncorrelated state and from a symmetric Kondo regime to a mixed valence regime. 
We find also for these quenches that the long-time limit of the spectral function approaches the expected one in the equilibrium final state upon increasing the number of quenches, with 32 quenches sufficing to obtain a similar accuracy as for
the quenches in Figs.~\ref{fig:infinite}(a) and \ref{fig:infinite}(b).

Clearly, the quality of the TDNRG spectral functions at long times, a key input within the scattering states NRG \cite{Anders2008a,Anders2008b}, can
be much improved by replacing the single-quench TDNRG in Refs.~\onlinecite{Anders2008a,Anders2008b} by the present multiple-
quench TDNRG. The use of the latter for this purpose should allow, in the future, for an accurate study of nonequilibrium steady states for bias voltages on scales of order at least $T_{\rm K}$. Furthermore, the high accuracy with which $Tr[\rho^{i\to f}(\tilde{\tau}_n)]=1$ 
is satisfied in the present formalism (see inset to Fig.~\ref{fig:compare2old-b}), guarantees that the spectral sum rule $\int d\omega A(\omega)=1$ is satisfied to a 
correspondingly high accuracy.

\subsection{Application to the long-time limit of thermodynamic observables}
\label{subsec:static-observables}

For further insight into the multiple quench TDNRG results, we also look at the results for thermodynamic observables in the long-time limit at finite temperatures. The percentage errors of the occupation number and the double occupancy  in the long-time limit when the system is switched from the mixed valence regime to the symmetric Kondo regime are shown in Figs.~\ref{fig:static}(a) and \ref{fig:static}(b), while the errors in the case of the reverse switching, i.e., from the symmetric Kondo regime to the mixed valence regime, are shown in Figs.~\ref{fig:static}(c) and \ref{fig:static}(d).
The percentage error is defined by the relative difference between the expectation value of the local observable in the long-time limit and the expected thermodynamic value in the final state, defined and denoted by $\delta O(t\to +\infty)=100\times\frac{O(t\to +\infty)-O_f}{O_f}$. 
In Fig.~\ref{fig:static} (a), the error of the occupation number in the case of a single quench is finite with an extremum at high temperature, and disappears only at the very highest temperature, $T>D$. With a larger number of quenches, $2$ and $8$, the absolute value of the error significantly decreases at low temperatures $T\leq T_{\rm K}$, the extrema also decrease in magnitude and remain at around the same temperature as observed in the results for a single quench. In the case of $31$ quenches, the error at low temperatures is closer to $0$ than in the other cases, and the extremum is also smaller but still finite. 
In Fig.~\ref{fig:static}(b), the error of the double occupancy in the case of single quench is positive at low temperatures $T\alt T_{\rm K}$ and negative at higher temperatures.  With an increasing number of quenches, the magnitude of the error at $T\alt T_{\rm K}$ is significantly reduced, approaching $0$, while the error around the high temperature extremum changes less significantly, and converges to a finite value with increasing number of quenches.
In the case of the reverse switching, Figs.~\ref{fig:static}(c) and \ref{fig:static}(d), the side shoulders at temperatures in the range of $7T_{\rm K}-40T_{\rm K}$ are also observed in addition to the extrema at higher temperature. 
With an increasing number of quenches, the errors decrease at low temperatures $T\leq T_{\rm K}$, and the errors around the high-temperature peaks also decrease but still remain finite.
The dependence of the error in the trace of the projected density matrices on the logarithmic discretization parameter $\Lambda$ is discussed in Appendix~\ref{sec:lambda-dependence}. The main finding there is that the error decreases with increasing $\Lambda$ for a sufficiently large number of quenches.

As mentioned in our previous paper \cite{Nghiem2014a}, the error in the long-time limit not only depends on the size of the quench but also on the largest incoherent excitation of the final state, $\varepsilon_{inc}^{max}=\max(|\varepsilon_f|,|\varepsilon_f+U_f|, \Gamma)$. Apparently, the TDNRG calculation for multiple quenches may overcome the first problem of quench size by dividing it into a sequence of smaller ones, but not the second problem since $\varepsilon_{inc}^{max}$ is the same in calculations for both single quench and multiple quenches. It suggests that the observed extrema at finite temperature may originate from the incoherent excitations. 

These results suggest that the TDNRG calculation for multiple quenches systematically improves the long-time limit of observables in the low-temperature regime $T\leq T_{\rm K}$, but not in the high-temperature regime for temperatures of order the scale of the highest-energy incoherent excitation. 
Nevertheless, the TDNRG presented is  promising for the study of the Kondo effect out of equilibrium, where the interest is primarily on low temperatures where a Kondo effect is present, and on the observed 
destruction of the  Kondo resonance when the bias voltage  is increased to values comparable to and above $T_{\rm K}$.

\subsection{TDNRG vs Exact diagonalization}
\label{subsec:static-observables-ed}
 
Finally, we apply the TDNRG formalism for multiple quenches with infinite switch-on time to the resonant level model, i.e., the Anderson impurity model with $U=0$, and compare the results to those of the ED study.

In the ED calculations, the conduction band is also discretized logarithmically using the parameter $\Lambda$ as in the NRG calculations, and the resulting model is likewise mapped onto an impurity coupled to a semi-infinite chain. The ED is applied to finite size initial and final state Hamiltonians of length $N$, corresponding to the longest chain diagonalized within a TDNRG approach\cite{Guettge2013b},  and one can then determine from the resulting single particle levels and eigenstates the time evolution of observables following a quench. We have generalized the formulas for the time evolution of observables within this approach, to the case of multiple quenches, and for more details we refer the reader to Appendix~\ref{sec:ExD}.  In the ED calculation, there is no truncation of states as in the NRG calculation, and one can therefore obtain approximation-free results (no  NRG approximation enters). The method can not be applied to the Anderson impurity model with $U\neq 0$, however. Since it also solves the same discrete model as in 
TDNRG, it can be used  as a benchmark to check the TDNRG calculations \cite{Guettge2013,Guettge2013b}. In addition, it can be formulated  for  infinite switch-on times (Appendix~\ref{sec:ExD}); this allows us to verify that an infinite switch-on time improves the
long-time limit of thermodynamic observables, as in the present multiple quench TDNRG approach. 

\begin{figure}[t]
  \centering
  \includegraphics[width=0.5\textwidth]{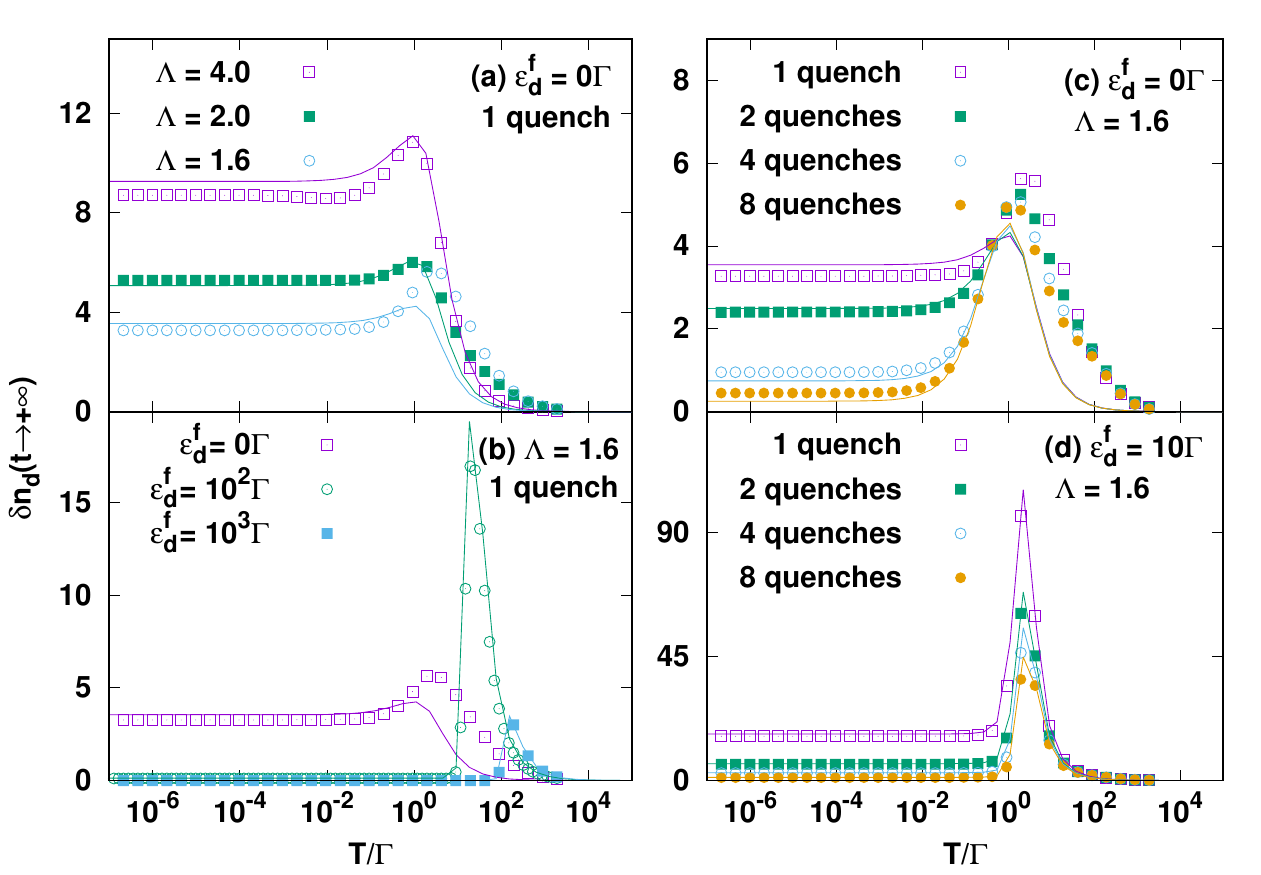}
    \caption{{The percentage error in the expectation value of the occupation number in the long-time limit, $\delta n_{d}(t\to\infty)$, vs the rescaled temperature, $T/\Gamma$, calculated by the TDNRG (symbols) and the exact diagonalization approach (solid lines) applied to the resonant level model. (a) Dependence of the single-quench results on $\Lambda$ for fixed final state level position $\varepsilon_d^f=0$ and a fixed quench size, $\Delta \varepsilon_d=\varepsilon_d^f-\varepsilon_d^i=6\Gamma$. (b) Dependence of the single-quench results on the final state level position $\varepsilon_d^f$ for  $\Lambda=1.6$ and a fixed quench size $\Delta \varepsilon_d=6\Gamma$. (c) and (d) show the dependence of the results on the number of quenches for two final state level positions: (c) $\varepsilon_d^f=0\Gamma$, and, (d), $\varepsilon_d^f=10\Gamma$, with fixed $\Lambda=1.6$ and fixed quench size $\Delta\varepsilon_d=6\Gamma$. NRG parameters: $N_s=900$ kept states per NRG iteration, and $N_z=8$ values for the $z$-averaging.
The TDNRG results are normalized by the numerically calculated trace of the projected density matrix at each temperature. For 
typical errors in the latter, see Appendix~\ref{sec:lambda-dependence}}.
    }\label{fig:IRLM}
\end{figure}

In Fig.~\ref{fig:IRLM}, we show the percentage error of the occupation number in the long-time limit calculated by both TDNRG and ED. 
Clearly, the TDNRG results almost overlap with the ED results. The difference is primarily visible at high temperatures and originates from the use of the truncation in the TDNRG (and absent in ED).
The ED calculation for a single quench in Figs.~\ref{fig:IRLM}(a) and \ref{fig:IRLM}(b) exhibits the same problem as in the TDNRG calculation, i.e., even at low temperature, where both methods yield largely the same result, this long-time result exhibits a 
finite ``error''. More precisely, this is largely not an error as such, but represents a deviation from the expected value for a continuum bath due to the use of a  logarithmically discretized bath. The latter is known to prevent perfect thermalization of observables to their expected values at long times within the single-quench TDNRG approach\cite{Rosch2012,Guettge2013,Nghiem2017}. The percentage error, at low temperature, is thus finite in both methods at low temperature and it shows an extremum at high temperature in both methods. 
Any remaining difference between the TDNRG and ED results can be attributed to truncation errors in the TDNRG approach (which can be seen to be small). Thus, the logarithmically discretized bath, and the consequent imperfect thermalization, 
is the main source of ``error'' in the long-time limit of observables. Arguably this imperfect thermalization should not be termed an ``error''  of the single-quench TDNRG approach, but a feature of this approach. 
With decreasing $\Lambda$ in Fig.~\ref{fig:IRLM} (a), i.e., better approximating the continuum bath, the error decreases significantly, and the extremum is still located at around the same high temperature. By changing $\varepsilon_d^f$ in Fig. \ref{fig:IRLM} (b), we can determine the relationship between the extremum at high temperature and the incoherent excitations, as defined above in Sec.~\ref{subsec:static-observables}. For example, when $\varepsilon_d^f=0\Gamma$, then $\varepsilon_{inc}^{max}=\Gamma$, we have that the corresponding extremum appears exactly at $T=\Gamma$. With larger $\varepsilon_d^f$, we have $\varepsilon_{inc}^{max}=\varepsilon_d^f$, and the extremum appears at higher temperatures around $\varepsilon_d^f$, but not exactly, due to the interference with the lower energy scale $\Gamma$. 

Turning now to the TDNRG and ED results for multiple quenches, we show in Figs.~\ref{fig:IRLM}(c) and \ref{fig:IRLM}(d) the percentage error in the long-time limit of the occupation number for
two equal sized quenches with two different values of $\varepsilon_d^f$ and in the limit of an infinite switch-on time. With increasing number of quenches, the errors in both cases are reduced close to $0$ at low temperature 
but the errors around the high-temperature extremum at $T\approx \varepsilon_{inc}^{max}=\varepsilon_d^f$ are always finite. In Fig.~\ref{fig:IRLM}(c), $\varepsilon_{inc}^{max}$ equals the lowest energy scale of the final system; then the extrema in the results with different number of quenches are almost the same. 

In summary, the ED calculations for both single and multiple quenches without any approximation also show errors in the occupation numbers in the long-time limit, with extrema at high temperatures as in TDNRG. These are largely due to the imperfect
thermalization in the long-time limit due to the use of a logarithmically discretized bath.
Dividing a large quench into a sequence of smaller ones with an infinite switch-on time, implemented in the ED calculations presented here, also improves the long-time limit of observables at low temperatures as in TDNRG. Any remaining small difference between the ED and TDNRG results is due to the use of truncation in the latter (absent in the former). This further supports the precision of the multiple-quench TDNRG results presented here for infinite switch-on times.

\section{Conclusions}\label{sec:conclusion}

In this paper, we developed an alternative multiple-quench TDNRG  formalism for general pulses, which reduces further the effect of the NRG approximation on the time evolution of observables. We showed this 
by comparison with the previous approach \cite{Nghiem2014a,Nghiem2014b}. Specifically, the trace of the projected density matrix versus time remains closer to 1 and the discontinuities in the time evolution of observables following 
quenches are significantly reduced. Both approaches improve the long-time limit of observables for increasing switch-on times, i.e., with increasing adiabaticity of the switching from initial to final state. However, the present approach
shows a monotonic and faster convergence of the long-time limit with increasing switch-on time than the previous approach. 
Moreover, the present formalism allows the limit of infinite switch-on time to be straightforwardly taken
analytically, which is impossible in the previous formalism. 

We also formulated the spectral function within the alternative formalism, both for finite and infinite switch-on times. For infinite switch-on time, we showed that the long-time limit of the zero-temperature spectral function approached its value in the equilibrium final state with high accuracy: the Friedel sum rule was satisfied to  within a few percent, which is to be compared with the much larger error 
of order typically 15\% in the single-quench approach \cite{Anders2008b,Nghiem2017}. 
Additional features,  at $|\omega|\lesssim T_{\rm K}$, found in the single-quench approach \cite{Nghiem2017}, are absent in the present approach. Hence, the present approach yields accurate 
results for the long-time limit of spectral functions for systems switched between an arbitrary initial and an arbitrary final state, overcoming the problems encountered within the single quench approach \cite{Rosch2012,Nghiem2017}. This improvement is particularly important for an accurate description of nonequilibrium steady states 
of quantum impurity systems, since methods such as the scattering states NRG approach \cite{Anders2008a} for addressing steady states, rely on an accurate time-evolved spectral function in the long-time limit.  In the future, we therefore plan to use the present multiple quench formalism to address nonequilibirum steady states in quantum impurity systems and to compare with known exact results \cite{Oguri2018,*Oguri2018Erratum1,Oguri2018b,Filippone2018} and other approaches \cite{Anders2008a,Cohen2014a,Dorda2015}. 

\begin{acknowledgments}
H. T. M. N acknowledges the support by Vietnam National Foundation for Science and Technology Development (NAFOSTED) under Grant No. 103.2-2017.353.
We acknowledge support by the Deutsche Forschungsgemeinschaft via the ``Research Training Group 1995'' and supercomputer support by the John von Neumann institute for Computing (J\"ulich).
\end{acknowledgments}

\appendix
\begin{widetext}

\section{Spectral function in the long-time limit}
\label{sec:spectra}

In order to evaluate the spectral function, we require an expression for the retarded two-time Green function $G_{BC}(t+t',t)=-i\theta(t')\operatorname{Tr}\{\hat{\rho}[\hat{B}(t+t'),\hat{C}(t)]_s\}$ within TDNRG. We work within the complete basis set and full density matrix approach \cite{Hofstetter2000,Peters2006,Weichselbaum2007}. Since we are here only interested in the long-time limit after the last quench, $t+t'>t>\tilde{\tau}_n$, we can write
\begin{align}
G_{BC}(t+t',t)&=-i\theta(t')\operatorname{Tr}\{\hat{\rho}[\hat{B}(t+t'),\hat{C}(t)]_s\}\nonumber\\
&=-i\theta(t')\operatorname{Tr}\{\hat{\rho}[e^{iH^{{f}}(t+t'-{\tilde\tau}_n)}...e^{iH^{Q_1}{\tau}_1}\hat{B}e^{-iH^{Q_1}{\tau}_1}...e^{-iH^{{f}}(t+t'-{\tilde\tau}_n)},e^{iH^{{f}}(t-{\tilde\tau}_n)}...e^{iH^{Q_1}{\tau}_1}\hat{C}e^{-iH^{Q_1}{\tau}_1}...e^{-iH^{{f}}(t-{\tilde\tau}_n)}]_s\}\nonumber\\
&=-i\theta(t')\operatorname{Tr}\{e^{-iH^f(t-{\tilde\tau}_n)}...e^{-iH^{Q_1}{\tau}_1}\hat{\rho}e^{iH^{Q_1}{\tau}_1}...e^{iH_f(t-{\tilde\tau}_n)}[e^{iH^ft'}\hat{B}e^{-iH^ft'},\hat{C}]_s\}
\end{align}
Inserting decompositions of unity $1=\sum_{lem}|lem\rangle \langle lem|$ in the above gives,
\begin{align}
G_{BC}(t+t',t)=&-i\theta(t')\sum_{\substack{lem\\l'e'm'\\l''e''m''}}\sum_{\substack{l_1e_1m_1\\l'_1e'_1m'_1}}...\sum_{\substack{l_ne_nm_n\\l'_ne'_nm'_n}}
{_f}\langle lem|e^{-iH^f(t-{\tilde\tau}_n)}e^{-iH^{Q_n}{\tau}_n}|l_{n}e_{n}m_{n}\rangle_{Q_{n}}...{_{Q_2}}\langle l_2e_2m_2|e^{-iH^{Q_1}{\tau}_1}|l_1e_1m_1\rangle_{Q_1}\nonumber\\
\times&{_{Q_1}}\langle l_1e_1m_1|\hat{\rho}|l'_1e'_1m'_1\rangle_{Q_1}{_{Q_1}}\langle l'_1e'_1m'_1|e^{iH^{Q_1}{\tau}_1}|l'_2e'_2m'_2\rangle_{Q_2}...{_{Q_n}}\langle l'_ne'_nm'_n|e^{iH^{Q_n}{\tau}_n}e^{iH_f(t-{\tilde\tau}_n)}|l'e'm'\rangle_{f}\nonumber\\
\times&({_f}\langle l'e'm'|e^{iH^ft'}\hat{B}e^{-iH^ft'}|l''e''m''\rangle_f{_f}\langle l''e''m''|\hat{C}|lem\rangle_f+{_f}\langle l'e'm'|\hat{C}|l''e''m''\rangle_f{_f}\langle l''e''m''|e^{iH^ft'}\hat{B}e^{-iH^ft'}|lem\rangle_f)
\label{eq:Gre}
\end{align}
Converting the multiple-shell summations over discarded states into a shell-diagonal restricted sum \cite{Note1} leads to
\begin{align}
G_{BC}(t+t',t)=-i\theta(t')\sum_m\sum_{rsqr_1s_1...r_ns_n}^{\notin KK'K''K_1K_1'...K_nK_n'}&S_{rr_n}^m
...S_{r_2r_1}^m\rho^{i\to Q_1}_m(r_1,s_1)e^{-i(E^m_{r_1}-E^m_{s_1}){\tau}_1}S^m_{s_1s_2}...e^{-i(E^m_{r_n}-E^m_{s_n}){\tau}_n}S^m_{s_ns}e^{-i(E^m_r-E^m_s)(t-\tilde{\tau}_n)}\nonumber\\
\times&({B}^m_{sq}e^{i(E^m_s-E^m_q)t'}C^m_{qr}+C^m_{sq}{B}^m_{qr}e^{i(E^m_q-E^m_r)t'}).
\end{align}
Fourier transforming this Green function with respect to the time difference $t'$ results in
\begin{align}
G_{BC}(\omega,t)=\sum_m\sum_{rsqr_1s_1...r_ns_n}^{\notin KK'K''K_1K_1'...K_nK_n'}&S_{rr_n}^m
...S_{r_2r_1}^m\rho^{i\to Q_1}_m(r_1,s_1)e^{-i(E^m_{r_1}-E^m_{s_1}){\tau}_1}S^m_{s_1s_2}...e^{-i(E^m_{r_n}-E^m_{s_n}){\tau}_n}S^m_{s_ns}e^{-i(E^m_r-E^m_s)(t-\tilde{\tau}_n)}\nonumber\\
\times&\Big(\frac{{B}^m_{sq}C^m_{qr}}{\omega+E^m_s-E^m_q+i\eta}+\frac{C^m_{sq}{B}^m_{qr}}{\omega+E^m_q-E^m_r+i\eta}\Big).
\end{align}
{
Then we have the spectral function, $A(\omega,t)=-Im[G(\omega,t)]/\pi$, in the long-time limit $t-\tilde{\tau}_n\to +\infty$ {for a finite switch-on time, i.e., for finite ${\tau}_{p=1,\dots,n}$,}
\begin{align}
A(\omega,t\to\infty)=
\sum_m\sum_{rqr_1s_1...r_ns_n}^{\notin KK'K_1K_1'...K_nK_n'}&S_{rr_n}^m
...S_{r_2r_1}^m\rho^{i\to Q_1}_m(r_1,s_1)e^{-i(E^m_{r_1}-E^m_{s_1}){\tau}_1}S^m_{s_1s_2}...e^{-i(E^m_{r_n}-E^m_{s_n}){\tau}_n}S^m_{s_nr}\nonumber\\
\times&
  \Big[{B}^m_{rq}C^m_{qr}\delta({\omega+E^m_r-E^m_q})+C^m_{sq}{B}^m_{qr}\delta({\omega+E^m_q-E^m_r})\Big],
\end{align}
and the long-time limit {for an infinite switch-on time, i.e., ${\tau}_{p=1,\dots,n}\to +\infty$,}
\begin{align}
A(\omega,t\to\infty)=\sum_m\sum_{rqr_1...r_n}^{\notin KK'K_1...K_n}&S_{rr_n}^m
...S_{r_2r_1}^m\rho^{i\to Q_1}_m(r_1,r_1)S^m_{r_1r_2}...S^m_{r_nr}\Big[{B}^m_{rq}C^m_{qr}\delta({\omega+E^m_r-E^m_q})+C^m_{sq}{B}^m_{qr}\delta({\omega+E^m_q-E^m_r})\Big],
\end{align}
  }
\end{widetext}
{where the restrictions $r_1=s_1,...,r_n=s_n$ and $r=s$ used in the above should more generally be replaced by $\delta_{E_{r_1},E_{s_1}},..., \delta_{E_{r_n},E_{s_n}}$ and  $\delta_{E_{r},E_{s}}$\footnote{See discussion in Sec.~\ref{sec:results}}.}
\section{Spectral function in the long-time limit: reverse quenches}
\label{sec:reverse-quenches}
\begin{figure}[ht]
  \centering 
  \includegraphics[width=0.5\textwidth]{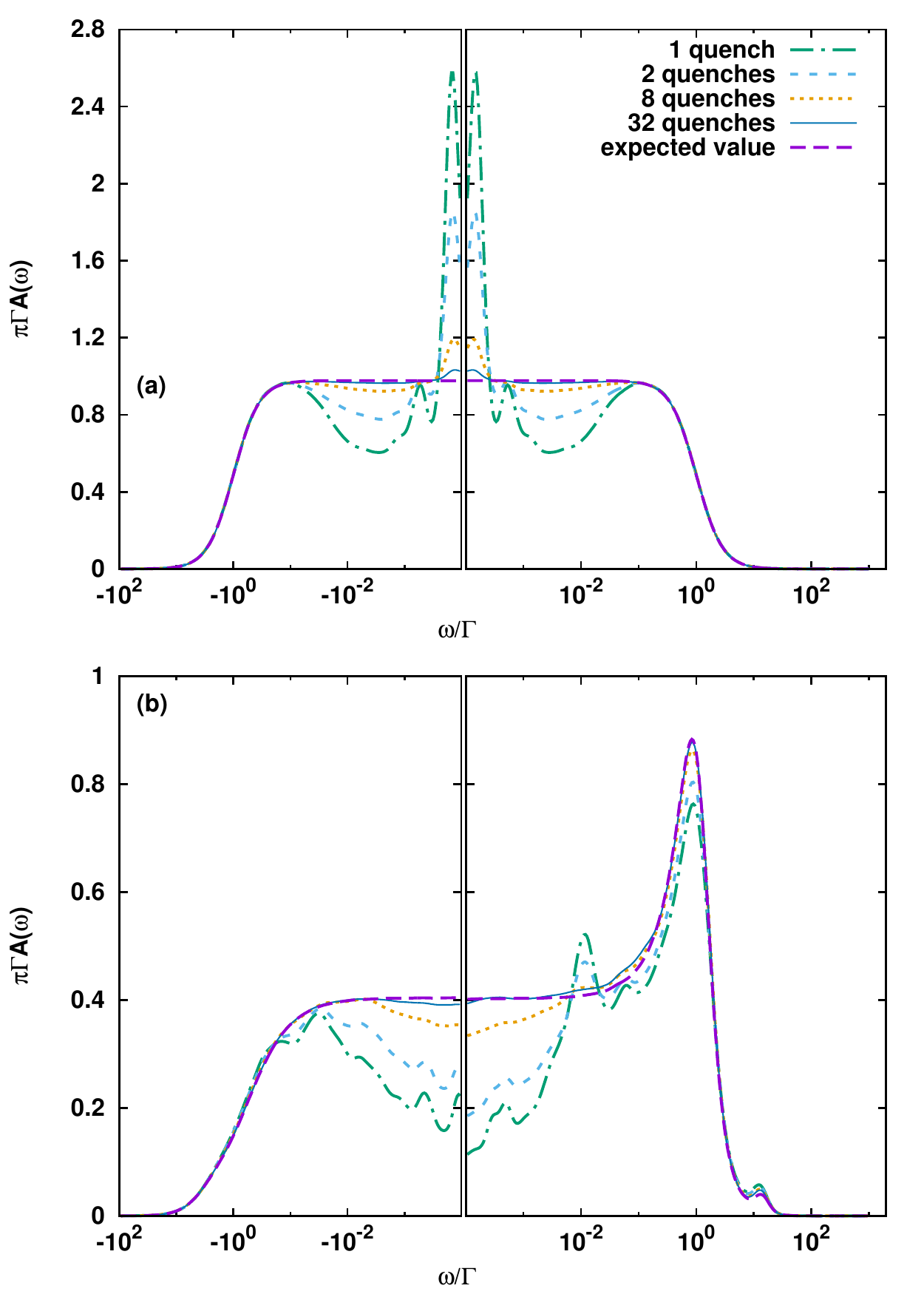}
\caption{
Normalized spectral function [$\pi\Gamma A(\omega)$] vs normalized frequency $\omega/\Gamma$ in the long-time limit and infinite switch-on time for $2$, $8$, and $32$ quenches 
compared to that from the single-quench TDNRG. Also shown is the expected value of the spectral function in the equilibrium final state. 
 (a) Switching from an interacting system with $\varepsilon_{d}^i=-U^i/2, U^i=12\Gamma$ to a noninteracting system with $\varepsilon_d^f=U^f=0$. 
(b) Switching from a system in the symmetric Kondo regime with $\varepsilon_{d}^i=-U^i/2, U^i=12\Gamma$ to one in the mixed valence regime with $\varepsilon_d^f=0$ and $U^f=12\Gamma$. 
$\Gamma=10^{-3}D$, and $D=1$ is the half-bandwidth. Calculations were for essentially zero temperature $T=10^{-4}T_{\rm K}$, with $T_{\rm K}$ the Kondo temperature in the initial state, 
NRG parameters: $\Lambda= 4$,  $E_{\text{cut}}=24$, and $N_z=8$ values for the $z$-averaging.
}
\label{fig:reverse}
\end{figure}
We show in Figs.\ref{fig:reverse}(a) and \ref{fig:reverse}(b) the long-time limit of the spectral function $A(\omega)=A(\omega,t\to\infty)$ for $1$, $2$, $8$, and $32$ quenches for the reverse of the two quenches in
Figs.~\ref{fig:infinite}(a) and \ref{fig:infinite}(b). While $A(\omega)$ exhibits significant substructures are low energies for a small number of quenches, these substructures are rapidly suppressed upon increasing
the number of quenches. For $32$ quenches, we recover in both cases the expected equilibrium spectral function of the final state to high accuracy. Thus, the Friedel sum rule in Fig.~\ref{fig:reverse}(a) is recovered
to within 5\%, while for the quench into the mixed valence regime in Fig.~\ref{fig:reverse}(b) it is recovered to within 3\%. In the latter, the mixed valence resonance is correctly renormalized upwards
from its bare value at $\omega=\varepsilon_d^f=0$ to $\omega=\tilde{\varepsilon}_d^f\approx \Gamma$ by the Coulomb interaction, while the higher lying satellite peak is also correctly located at $\omega\approx \varepsilon_d^f+U^f\approx 12\Gamma$\cite{Costi1994}. Similarly, for the quench 
in Fig.~\ref{fig:reverse}(a) we see that $A(\omega)$ for 32 quenches recovers the noninteracting resonant level of the final state with halfwidth at half maximum given by $\Gamma$ and centered at zero energy.
In conclusion, for a sufficient number of quenches, the present formalism for an infinite switch-on time is able to describe the long-time limit of the spectral function to high accuracy. In the next section, we discuss the
effects of a finite switch-on time on $A(\omega)$ in the long-time limit.

\section{Spectral function in the long-time limit: dependence on a finite switch-on time}
\label{sec:finiteS}
\begin{figure}[ht]
  \centering 
  \includegraphics[width=0.5\textwidth]{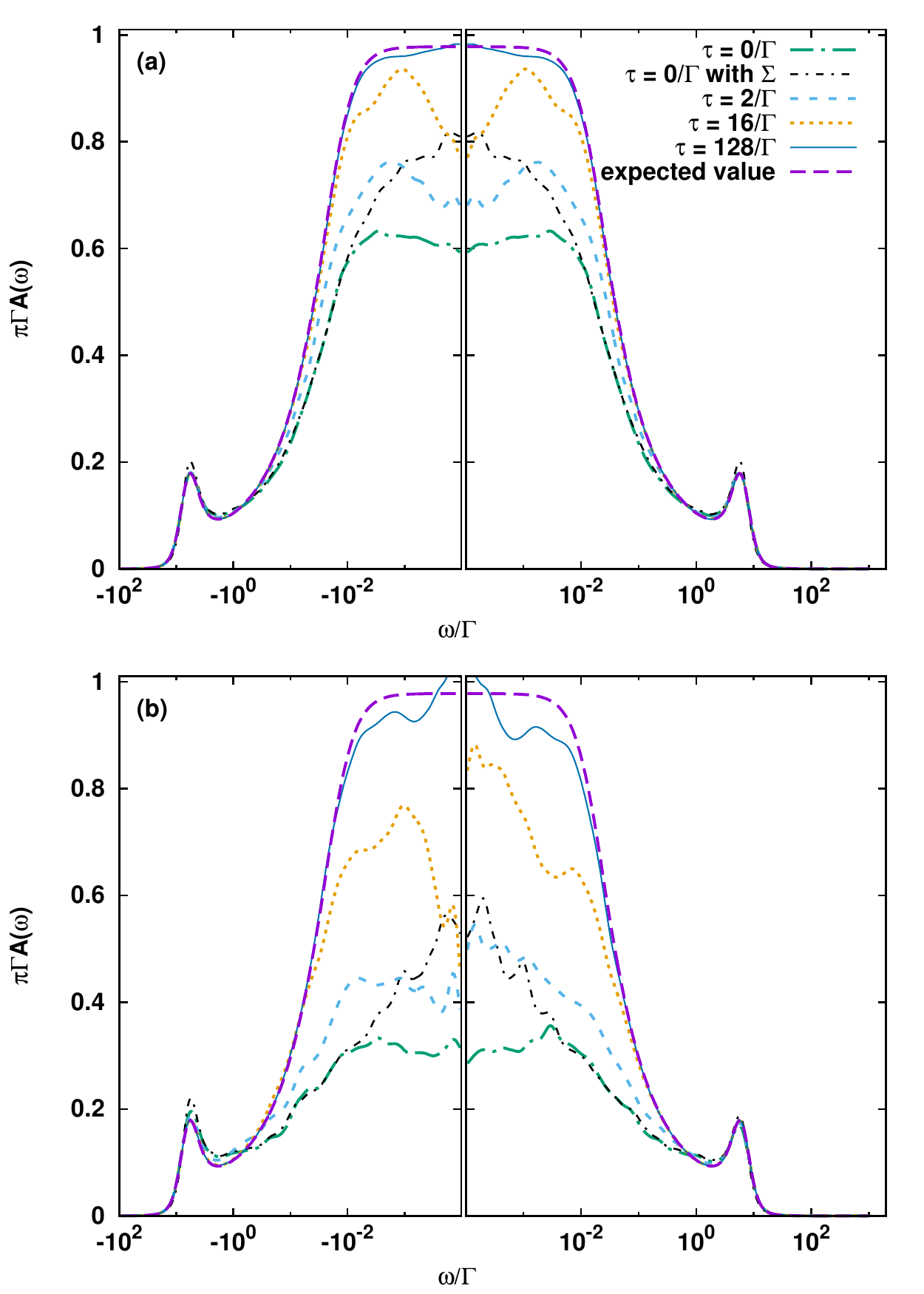}
\caption{Spectral function in the long-time limit $A(\omega,t\to\infty)$ vs $\omega/\Gamma$ for different finite switch-on times $\tau=\tilde{\tau}_n$. Also shown is the expected equilibrium spectral function in the final state. 
(a) Switching from a noninteracting regime with $\varepsilon_d^{i}=0$ and $U^{i}=0$ to an interacting Kondo regime with $\varepsilon_d^{f}=-U^{f}/2$ and $U^{f}=12\Gamma$.  
(b) Switching from the mixed valence regime with $\varepsilon_d^{i}= 0$ and $U^{i}=12\Gamma$ to the symmetric Kondo regime with $\varepsilon_d^{f}=-U^{f}/2$ and $U^{f}=12\Gamma$. 
$\Gamma=10^{-3}D$, with $D=1$ the half-bandwidth. Calculations were for essentially zero temperature, $T=10^{-4}T_{\rm K}$, with $T_{\rm K}$ the 
Kondo temperature in the initial state. 
NRG parameters: $\Lambda= 4$,  $E_{\text{cut}}=24$, and $N_z=8$ values for the $z$-averaging. 
}\label{fig:finite}
\end{figure}

We show in Fig.~\ref{fig:finite} the dependence of the long-time limit of the spectral function $A(\omega)=A(\omega,t\to\infty)$ on the switch-on time $\tau=\tilde{\tau}_n$. As with the occupation number, the long-time limit of the spectral function also improves and approaches the expected value in the equilibrium final state with increasing switch-on time $\tau$. In the case of switching from the asymmetric to symmetric Kondo regime [Fig.~\ref{fig:finite}(b)], the spectral function in the long-time limit becomes more symmetric with increasing $\tau$. 
However,  the spectral function shows small additional structures at $|\omega|<T_{\rm K}$ even when the switch-on time $\tau$ exceeds the time scale $1/T_{\rm K}$ for the formation of the Kondo resonance\cite{Nordlander1999,Nghiem2017}. The error in the spectral sum rule $\int_{-\infty}^{+\infty}d\omega A(\omega,t\to\infty)=1$ is violated in this case by $0.1\%$. This is attributed to the NRG approximation in the multiple quench formalism,  which results in a cumulative error in the trace of the projected density matrix and a discontinuity in the time evolution of observables, as discussed in Sec.~\ref{sec:accuracy}. In the case of switching from a noninteracting to an interacting system,  Fig. \ref{fig:finite}(a), the long-time limit of the spectral function lies closer to the expected result than that for the second switching protocol in Fig. \ref{fig:finite}~(b) for each $\tau$. 
However, at the longest $\tau$, additional structures within the Kondo resonance at  $|\omega|<T_{\rm K}$ are still visible, attributable to the more pronounced effect of the NRG approximation in the case of finite switch-on times.
The above findings support the conclusion that accurate results can be obtained for the long-time limit 
of the spectral function within TDNRG by replacing a single large quench by a sequence of smaller quenches and switching the system slowly from one state to the other (i.e., with increasing $\tau$). The most accurate results are obtained
in the limit of a large number of quenches and for $\tau\to\infty$ as supported by the results in Sec.~\ref{subsec:spectral-function} and Appendix \ref{sec:reverse-quenches}.

\section{Calculations with different $\Lambda$}
\label{sec:lambda-dependence}
\begin{figure}[ht]
    \centering
      \includegraphics[width=0.5\textwidth]{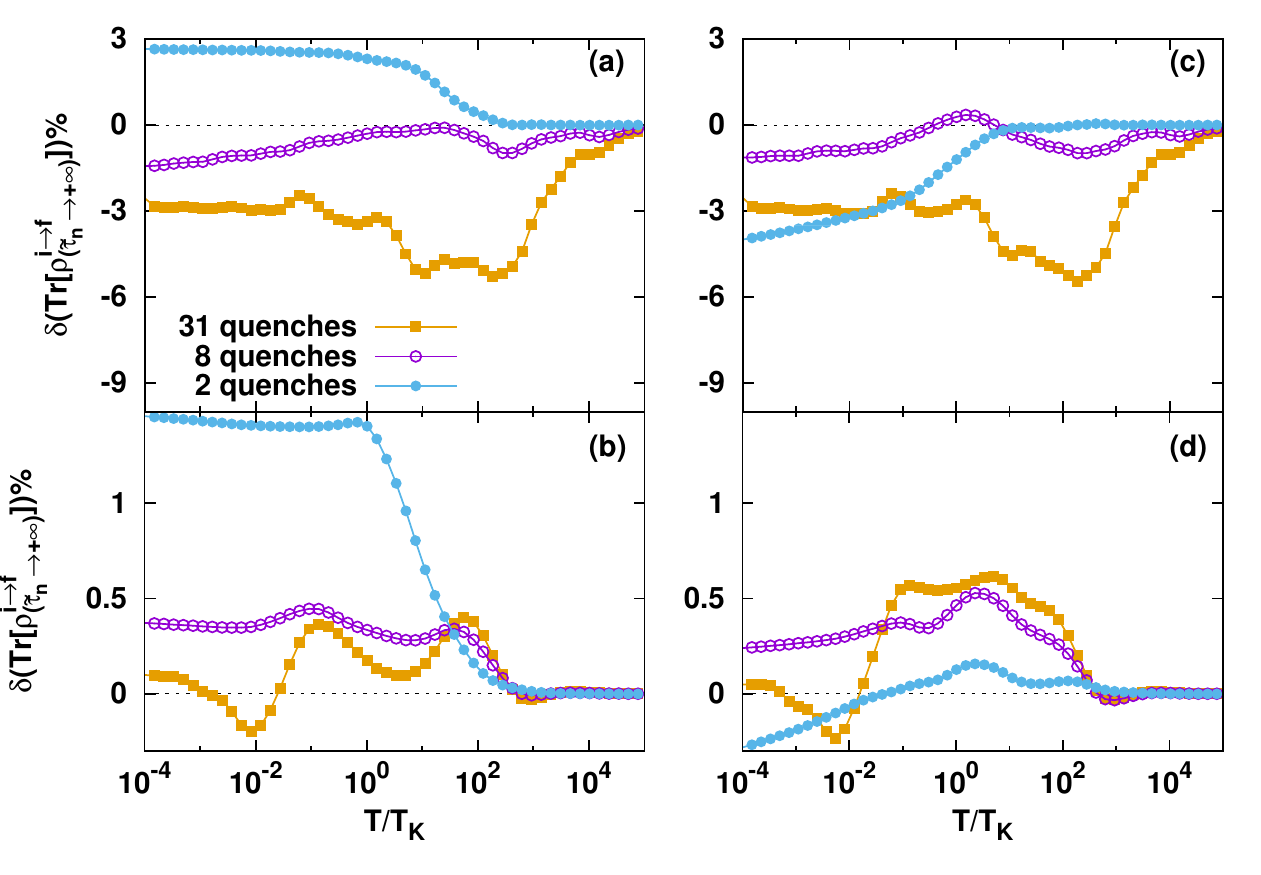}
      \caption{Percentage deviation of the trace of the projected density matrix away from $1$ for different $\Lambda$ vs rescaled temperature $T/T_{\rm K}$. Upper panels (a) and (c) are for $\Lambda=1.6$. Lower panels (b) and (d) are for $\Lambda=4$. 
        Left panels (a) and (b) were for switching from the mixed valence regime with $\varepsilon_d^{i}= 0$ and $U^{i}=12\Gamma$ to the symmetric Kondo regime with $\varepsilon_d^{f}=-U^{f}/2$ and $U^{f}=12\Gamma$, while
        the right panels (c) and (d) were for switching from the symmetric Kondo regime with $\varepsilon_d^{i}=-U^{i}/2$ and $U^{i}=12\Gamma$ to the mixed valence regime with $\varepsilon_d^{f}= 0$ and $U^{f}=12\Gamma$. 
        $\Gamma=10^{-3}D$, with $D=1$ the half-bandwidth. $T_{\rm K}$ is the
        Kondo temperature of the symmetric Kondo state. NRG parameters for the calculations with $\Lambda=1.6$: $N_z=8$ and the number of kept states $N_s=900$. For the calculations with $\Lambda=4$, the parameters are the same as those in Fig.~\ref{fig:static}.
      }\label{fig:rho}
    \end{figure}

We show in Fig.~\ref{fig:rho} the errors of the traces of the projected density matrices versus temperature for calculations with different $\Lambda$. 
In the calculations with $\Lambda=1.6$ in Figs.~\ref{fig:rho}(a) and \ref{fig:rho}(c),  the percentage errors are as large as approximately $3\%$ at low temperatures, and they exhibit an extremum of up to approximately $6\%$ at high temperatures. 
The absolute values of the errors are non-monotonic with respect to the number of quenches.
For the case of $31$ quenches, the errors in both quenches are similar, suggesting that for the large number of quenches the error strongly depends on the quench size. 
The calculations with $\Lambda=4$ [Fig.~\ref{fig:rho}(b) and Fig.~\ref{fig:rho}(d)] result in much smaller errors than for $\Lambda=1.6$. 
Except for the error in the case of two quenches in Fig.~\ref{fig:rho} (b) which is up to around $1.5\%$, all the errors for a larger number of quenches are less than $0.6\%$. 
We conclude that the formalism presented here results in smaller errors in the trace of the projected density matrices with increasing values of $\Lambda$.

\section{Exact diagonalization of the resonant level model with multiple quenches}
\label{sec:ExD}
The real-time revolution of a system, modeled by the RLM, following a single quench can be calculated via 
ED \cite{Guettge2013b}. In this appendix, we derive results for the time dependence of the occupation number 
of the resonant level model, $\langle n_d(t)\rangle$, and also for the time dependence of the conduction electron orbital occupation numbers, first for the case of two quenches, and then generalizing this to the case of an arbitrary number of quenches. The presented expressions are then free of any approximations, both for finite and infinite switch-on times.

In the ED of the RLM, the conduction band is also discretized with the parameter $\Lambda$ and mapped onto a Wilson chain 
as in the NRG calculation. Then we have the following discrete model;
\begin{align}
H_N(t)&=\varepsilon_d(t)d^{\dagger}d+V(t)(d^{\dagger}c_{0}+c^{\dagger}_{0}d)+\sum_{n=0}^{N-2}t_n(c^{\dagger}_nc_{n+1}+c^{\dagger}_{n+1}c_n)\nonumber\\
&=\vec{\alpha}^{\dagger}M(t)\vec{\alpha},
\end{align}
with $\vec{\alpha}=\left( \begin{array}{c} d \\c_0 \\\vdotswithin{}\end{array} \right)$, $\vec{\alpha}^{\dagger}=\left( \begin{array}{ccc} d^{\dagger} & c_0^{\dagger} &\cdots\end{array} \right)$, and
\begin{equation}
M(t)=\left( \begin{array}{cccc} \varepsilon_d(t) & V & 0 & 0\\ V & 0 & t_0 & 0\\ 0 & t_0 & 0 &\ddots\\ 0 & 0 &\ddots &\ddots\end{array} \right).
\end{equation}
$M(t)$ can be diagonalized as follows
\begin{equation}
M(t)=U(t)^{\dagger}\text{diag}(\epsilon_1(t),\epsilon_2(t),\cdots)U(t).
\end{equation}
\begin{figure}[ht]
\centering 
\includegraphics[width=0.29\textwidth]{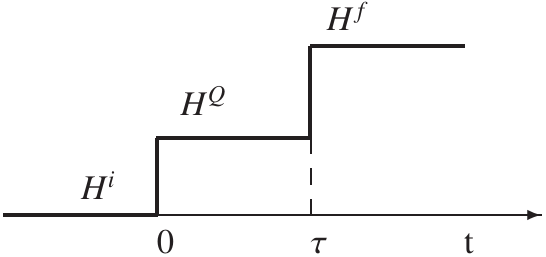}
\caption 
{A system driven from the initial $H^{i}$ to the final state $H^{f}$ via the intermediate state described by $\{H^{Q}\}$.
}
\label{fig:3steps}
\end{figure}
$H(t)$ is represented in Fig. \ref{fig:3steps}, in which each Hamiltonian can be expressed in the diagonal form as
$H^i=\sum_n\epsilon_n^if^{i\dagger}_nf^i_n$, $H^Q=\sum_n\epsilon_n^Qf^{Q\dagger}_nf^Q_n$, and $H^f=\sum_n\epsilon_n^ff^{f\dagger}_nf^f_n$ with
\begin{align}
&f^i_n=\sum_lU^i_{nl}\alpha_l,\quad f^Q_n=\sum_lU^Q_{nl}\alpha_l\quad, f^f_n=\sum_lU^f_{nl}\alpha_l\\
&f^{i\dagger}_n=\sum_l\alpha^{\dagger}_lU^{i\dagger}_{nl},\quad f^{Q\dagger}_n=\sum_l\alpha^{\dagger}_lU^{Q\dagger}_{nl},\quad f^{f\dagger}_n=\sum_l\alpha^{\dagger}_lU^{f\dagger}_{nl}.
\end{align}
The operator for the occupation number of site $m$ is defined as
\begin{equation}
n_m=\alpha^{\dagger}_m\alpha_m=
  \begin{dcases}
     d^{\dagger}d & m=0;\\
     c^{\dagger}_{m-1}c_{m-1} & m>1.
   \end{dcases}
\end{equation}
The expectation value of the occupation number is given by
\begin{equation}
\langle n_m(t>\tau)\rangle=Tr[\rho(t)n_m],
\end{equation}
with $\rho(t)=e^{-iH^f(t-\tau)}e^{-iH^Q\tau}\rho_0e^{iH^Q\tau}e^{iH^f(t-\tau)}$ and $\displaystyle\rho_0=\frac{e^{-\beta H^i}}{Z}=\frac{e^{-\beta\sum_n\epsilon^i_n f^{i\dagger}_nf^i_n}}{Tr[e^{-\beta\sum_n\epsilon^i_n f^{i\dagger}_nf^i_n}]}$. Then we have
\begin{align}
\langle n_m(t>\tau)\rangle&=Tr[e^{-iH^f(t-\tau)}e^{-iH^Q\tau}\rho_0e^{iH^Q\tau}e^{iH^f(t-\tau)}n_m]\nonumber\\
&=Tr[e^{-iH^Q\tau}\rho_0e^{iH^Q\tau}\underbrace{e^{iH^f(t-\tau)}n_me^{-iH^f(t-\tau)}}_{n_m(t-\tau)}].\label{eq:c8}
\end{align}
\begin{align}
n_m(t-\tau)&=e^{iH^f(t-\tau)}\underbrace{\alpha_m^{\dagger}\alpha_m}_{n_m}e^{-iH^f(t-\tau)}\nonumber\\
&=\sum_{nn'}e^{iH^f(t-\tau)}f^{f\dagger}_nf^f_{n'}e^{-iH^f(t-\tau)}U^f_{nm}U^{f\dagger}_{mn'}\nonumber\\
&=\sum_{nn'}\underbrace{e^{iH^f(t-\tau)}f^{f\dagger}_n e^{-iH^f(t-\tau)}}_{f^{f\dagger}_n(t-\tau)}\underbrace{e^{iH^f(t-\tau)}f^f_{n'} e^{-iH^f(t-\tau)}}_{f^{f}_{n'}(t-\tau)}U^f_{nm}U^{f\dagger}_{mn'}\nonumber\\
&=\sum_{nn'}e^{i(\epsilon^f_n-\epsilon^f_{n'})(t-\tau)}f^{f\dagger}_nf^f_{n'} U^f_{nm}U^{f\dagger}_{mn'},\label{eq:c9}
\end{align}
since $\partial_{(t-\tau)} f^{f}_{n}(t-\tau)=i[H^f,f^{f}_{n}(t-\tau)]=-i\epsilon^f_n f^{f}_{n}(t-\tau)$, then $f^{f}_{n}(t-\tau)=e^{-i\epsilon^f_n(t-\tau)}f^{f}_{n}$. Substituting (\ref{eq:c9}) into (\ref{eq:c8}), we have
\begin{align}
&\langle n_m(t>\tau)\rangle=\sum_{nn'}e^{i(\epsilon^f_n-\epsilon^f_{n'})(t-\tau)}Tr[e^{-iH^Q\tau}\rho_0e^{iH^Q\tau}f^{f\dagger}_nf^f_{n'}] U^f_{nm}U^{f\dagger}_{mn'}\nonumber\\
&=\sum_{nn'}e^{i(\epsilon^f_n-\epsilon^f_{n'})(t-\tau)}\sum_{kk'}Tr[e^{-iH^Q\tau}\rho_0e^{iH^Q\tau}f^{Q\dagger}_kf^Q_{k'}]\nonumber\\
&\hspace{2em}\times(U^QU^{f\dagger})_{kn}(U^fU^{Q\dagger})_{n'k'} U^f_{nm}U^{f\dagger}_{mn'}.\label{eq:c10}
\end{align}
Similarly, the trace in \ref{eq:c10} is evaluated as follows
\begin{align}
&Tr[e^{-iH^Q\tau}\rho_0e^{iH^Q\tau}f^{Q\dagger}_kf^Q_{k'}]\nonumber\\
=&Tr[\rho_0e^{iH^Q\tau}f^{Q\dagger}_ke^{-iH^Q\tau}e^{iH^Q\tau}f^Q_{k'}e^{-iH^Q\tau}]\nonumber\\
=&e^{i(\epsilon^Q_k-\epsilon^Q_{k'})\tau}Tr[\rho_0f^{Q\dagger}_kf^Q_{k'}]\nonumber\\
=&e^{i(\epsilon^Q_k-\epsilon^Q_{k'})\tau}\sum_{qq'}\underbrace{Tr[\rho_0f^{i\dagger}_qf^i_{q'}]}_{f(\epsilon_q^i)\delta_{qq'}}(U^iU^{Q\dagger})_{qk}(U^QU^{i\dagger})_{k'q'}\nonumber\\
=&e^{i(\epsilon^Q_k-\epsilon^Q_{k'})\tau}\sum_{q}f(\epsilon_q^i)(U^iU^{Q\dagger})_{qk}(U^QU^{i\dagger})_{k'q}\label{eq:c11},
\end{align}
in which $f(\epsilon_q^i)$ is the Fermi-Dirac distribution. Substituting \ref{eq:c11} into \ref{eq:c10}, we have
\begin{align}
&\langle n_m(t>\tau)\rangle\nonumber\\
&=\sum_{nn'}e^{i(\epsilon^f_n-\epsilon^f_{n'})(t-\tau)}\sum_{kk'}e^{i(\epsilon^Q_k-\epsilon^Q_{k'})\tau}\sum_{q}f(\epsilon_q^i)(U^iU^{Q\dagger})_{qk}(U^QU^{i\dagger})_{k'q}\nonumber\\
&\hspace{2em}\times(U^QU^{f\dagger})_{kn}(U^fU^{Q\dagger})_{n'k'} U^f_{nm}U^{f\dagger}_{mn'}.\label{eq:c12}
\end{align}
Defining 
\begin{align}
n^{i\to Q}_{kk'}&=\sum_{q}f(\epsilon_q^i)(U^iU^{Q\dagger})_{qk}(U^QU^{i\dagger})_{k'q}\label{eq:c13}\\
n^{i\to f}_{nn'}&=\sum_{kk'}e^{i(\epsilon^Q_k-\epsilon^Q_{k'})\tau}n^{i\to Q}_{kk'}(U^QU^{f\dagger})_{kn}(U^fU^{Q\dagger})_{n'k'},
\end{align}
we have 
\begin{align}
&\langle n_m(t>\tau)\rangle=\sum_{nn'}e^{i(\epsilon^f_n-\epsilon^f_{n'})(t-\tau)}n^{i\to f}_{nn'} U^f_{nm}U^{f\dagger}_{mn'}.\label{eq:c15}
\end{align}
These expressions are generalized to the case of $(p+1)$ quenches as follows
\begin{align}
n^{i\to f}_{nn'}&=\sum_{kk'}e^{i(\epsilon^{Q_p}_k-\epsilon^{Q_p}_{k'}){\tau}_p}n^{i\to Q_p}_{kk'}(U^{Q_p}U^{f\dagger})_{kn}(U^fU^{Q_p\dagger})_{n'k'},\label{eq:c16}
\end{align}
which is a recursion relation allowing $n^{i\to Q_p}$ to be  derived from $n^{i\to Q_{p-1}}$, and consequently from $n^{i\to Q_1}$ determined in \ref{eq:c13}. 
Finally, we have for the occupation of the orbitals
\begin{align}
&\langle n_m(t>\tilde{\tau}_p)\rangle=\sum_{nn'}e^{i(\epsilon^f_n-\epsilon^f_{n'})(t-\tilde{\tau}_p)}n^{i\to f}_{nn'} U^f_{nm}U^{f\dagger}_{mn'},\label{eq:c17}
\end{align}
with $\tilde{\tau}_p=\sum_{i=1}^p\tau_i$. 

The formulas above for the real-time dynamics following multiple quenches within ED is without any approximation. The extension  to the case of an infinite switch-on time, $\tilde{\tau}_p\to +\infty$, is obtained straight forwardly by setting $k=k'$ in \ref{eq:c16}, and
yields the time independent long-time limit result for the occupation numbers. 
\bibliography{noneq-nrg}
\end{document}